\parindent=0cm 

\font\gross=cmr10 scaled \magstep1

\font\mengen=bbm10

\font\csczwoelf=cmcsc10 scaled \magstep1

\font\ninebf=cmb7

\def\datum{\line{\hfil Berlin, den \the\day.\the\month.\the\year}}
\def\date{\line{\hfil Berlin,  \the\month/\the\day/\the\year}}

\def\doppel{\magnification=\magstep1}

\def\titel#1#2{{\removelastskip\bigskip\goodbreak\noindent\mark{#1}\gross
#1\medskip \nobreak\noindent #2\unskip}}
\def\subtitel#1#2{{\removelastskip\bigskip\goodbreak\noindent\bf
#1\medskip \nobreak\noindent #2\unskip}}
\edef\ignore#1{}

\def\lrkopfzeile{\nopagenumbers\headline={\ifodd\pageno\ungeraderkopf\else
\geraderkopf\fi}
\voffset=2\baselineskip}
\def\geraderkopf{\rm\folio\ \dotfill\ \it\firstmark}
\def\ungeraderkopf{\it\firstmark\ \dotfill\ \rm\folio}

\def\tr{\hbox{tr}}
\def\part#1#2{{\partial #1\over\partial #2}}
\def\br#1#2{{#1\over #2}}
\def\frac#1#2{{#1\over #2}}

\def\sla#1{\hbox to 0cm{/\hss}#1}
\def\MN{\hbox{\mengen N}}
\def\MZ{\hbox{\mengen Z}}
\def\MR{\hbox{\mengen R}}

\def\ID{\hbox{\mengen 1}}


\def\lrvec#1{\vbox{\ialign{##\crcr$\leftrightarrow$\crcr\noalign{\kern-1pt\nointerlineskip}$\hfil\displaystyle{#1}\hfil$\crcr}}}
\def\pfeilrmittext#1{\setbox3=\hbox{#1\kern
0.5em}$\displaystyle\mathrel{\mathop {\hbox to
\wd3{\rightarrowfill}}^{\box3}}$}  
\def\pfeillmittext#1{\setbox3=\hbox{\kern
0.5em #1}$\displaystyle\mathrel{\mathop {\hbox to
\wd3{\leftarrowfill}}^{\box3}}$}  
\def\pfeilrmittextou#1#2{\setbox3=\hbox{#1\kern
0.5em}$\displaystyle\mathrel{\mathop {\hbox to
\wd3{\rightarrowfill}}^{\box3}_{\hbox{#2\kern 0.5em}}}$}  
\def\pfeillmittextou#1#2{\setbox3=\hbox{\kern
0.5em #1}$\displaystyle\mathrel{\mathop {\hbox to
\wd3{\leftarrowfill}}^{\box3}_{\hbox{\kern 0.5em #2}}}$}  
%

%

\def\putlogo{\vbox to 0cm{\hbox to
0cm{\noindent\line{\hss\epsfbox{AEIlogo.eps}}\hss}\vss}}

\newcount\figcount \figcount=0
\def\fig#1#2{\global\advance\figcount by1\midinsert\vskip #2
\centerline{Fig.\ \the\figcount : #1}\endinsert}
\def\figstuff#1#2{\global\advance\figcount by1\midinsert #2
\centerline{Fig.\ \the\figcount : #1}\endinsert}
\def\nextfig{Fig.~{\advance\figcount by 1\relax\the\figcount}}
\def\psfig#1#2{\global\advance\figcount
by 1 \midinsert\vbox{\centerline{\epsfbox{#2}}
\centerline{Fig.\ \the\figcount : #1}}\endinsert}
\def\rpsfig#1#2#3{\global\advance\figcount
by 1 \xdef\#3{\the\figcount}\midinsert\vbox{\centerline{\epsfbox{#2}}
\centerline{Fig.\ \the\figcount : #1}}\endinsert} 

\def\ez#1e#2{#1\cdot 10^{#2}}
\def\crossout#1{\hbox to 0cm{\raise 1ex \hbox{$\underline{\hbox{\phantom{#1}}}$}\hss}#1}
\def\foryoureyesonly#1{}
%
\def\paper#1#2#3#4#5#6#7#8#9{\item{\bf [#1]}#2: ``#3'', #4 {\bf
#5} (#6) p.\ #7 #8 \foryoureyesonly{#9}\par\goodbreak}
%
%
\def\buch#1#2#3#4#5#6#7#8{\item{\bf [#1]}#2: ``#3'' (#4) #5 #6 #7 \foryoureyesonly{#8}\par}
%
%
\def\eprint#1#2#3#4#5#6{\item{\bf [#1]}#2: ``#3'', {{\tt #4}} #5
\foryoureyesonly{#6}\par}
%
%

%
%
\newcount\chapterno
\newcount\subchapterno
\newcount\glno
\edef\rememberref#1{\relax}  
\def\rememberchapter#1{\relax}  
\def\remembersubchapter#1{\relax}  
\def\neueseitevorchapter{\vfill\supereject\ifodd\pageno\relax\else \noindent$
$\vfill\eject\fi} 
\def\kapitelnummer{\the\chapterno}
\def\chapter#1{\neueseitevorchapter\global\advance\chapterno by
  1\expandafter\rememberchapter{\the\chapterno . #1}\glno =0
  \subchapterno =0\titel{\the\chapterno . #1}} 
\def\subchapter#1{\global\advance\subchapterno by
    1\remembersubchapter{\the\chapterno .\the\subchapterno
    . #1}\subtitel{\the\chapterno .\the\subchapterno . #1}}
\def\gln#1{\global\advance\glno by 1\xdef#1{(\the\chapterno
.\the\glno)}\eqno {(\the\chapterno .\the\glno)\ifdraft\hbox to 0cm{\tt\string #1\hss}\else\relax\fi}}
\def\egln#1{\global\advance\glno by 1\xdef#1{(\the\chapterno
.\the\glno)}& {(\the\chapterno .\the\glno)\ifdraft\hbox to 0cm{\tt\string #1\hss}\else\relax\fi}}
\newif\ifdraft
\draftfalse
\def\draft{\drafttrue\def\vielluft{\bigskip}\def\neueseitevorchapter{\par}\def\foryoureyesonly##1{##1}\def\shlabel##1{\hbox
  to 0cm{\tt \expandafter\string ##1\hss}}}
\def\shlabel#1{\relax} 

\def\openbib{\def\aux{1}\openout\aux=\jobname.aux \immediate\write16{Putting
    references in \jobname.aux}
    \def\rememberref##1{\write\aux{cite(##1) on page \folio}}
    \def\rememberchapter##1{{\edef\x{\noexpand\write\aux{chapter[##1] on page
    \noexpand\folio}}\x}}
    \def\remembersubchapter##1{\write\aux{subchapter[##1] on page \folio}} }
\def\closebib{\closeout\aux \def\rememberref##1{\relax}}
\def\cite#1{\raise 0.5ex\hbox{\ninebf[#1]}\rememberref{#1}}

\newif\ifpdf
\ifx\pdfoutput\undefined\pdffalse\else\pdfoutput=1\pdftrue\fi

{\catcode`\%=11\catcode`\!=14
\gdef\bluebg{\pdfliteral{
}
\def\aeiseminar{\magnification = 2500\raggedright
        \ifpdf\pdfpagewidth=35true cm\hsize=
        \pdfpagewidth  
        \bluebg
        \BrickRed\else\textBrickRed\fi}
\def\color{\ifpdf\input pdfcolor\else\input colordvi\fi}


 
\input epsf.tex
\def\neueseitevorchapter{\relax}
\def\mod{\mathop{\rm mod}\nolimits}
\doppel
\baselineskip=1.2\baselineskip
\openbib
\parindent=1true cm
{\nopagenumbers
\line{\hss HU-EP-02/11}
\line{\hss AEI-2002/026}
\line{\hss hep-th/0204037}
\vskip1cm
\centerline{\csczwoelf Perturbative Instabilities on the Non-Commutative Torus,}
\centerline{\csczwoelf Morita Duality and Twisted Boundary Conditions} 
\vskip1cm
{\it
\centerline{Zachary Guralnik$^\dagger$\footnote{$^1$}{{\tt zack@physik.hu-berlin.de}}, 
Robert C. Helling$^\dagger$\footnote{$^2$}{{\tt helling@AtDotDe.de}},
Karl Landsteiner$^\dagger$\footnote{$^3$}{{\tt Karl.Landsteiner@physik.hu-berlin.de}}}
\centerline{ and 
Esperanza Lopez$^\diamond$\footnote{$^4$}{{\tt lopez@aei-potsdam.mpg.de}}}
}
\vskip1cm
\centerline{$^\dagger$ Institut f\"ur Physik}
\centerline{Humboldt-Universit\"at zu Berlin, Invalidenstra\ss e 110}
\centerline{D-10115 Berlin, Germany}
\bigskip
\centerline{$^\diamond$ Max Planck Institut f\"ur Gravitationsphysik}
\centerline{Albert Einstein Institut, Am M\"uhlenberg 1}
\centerline{D-14476  Golm, Germany}
\vskip2cm 
\centerline{\bf Abstract:}
\noindent We study one-loop corrections in scalar and gauge field theories on the non-commutative
torus. For rational $\theta$, Morita equivalence
allows these theories to be reformulated in terms of ordinary theories on a
commutative torus with twisted boundary conditions. UV/IR mixing does not lead to 
singularities, however there can be large corrections.
In particular, gauge theories show tachyonic
instabilities for some of the modes. We discuss their relevance to spontaneous
$\MZ_N\times \MZ_N$ symmetry breaking in the Morita dual $SU(N)$ 
theory due to electric flux condensation.
\vskip 3cm plus 0.1fil

\eject}

\chapter{Introduction}
For a variety of reasons, non-commutative field theories have been of much
recent interest. One motivation for studying such theories is
that the notion of space-time
presumably has to be modified at very short distances \cite{Sn}\cite{C}\cite{DFR}.
Space-time non-commutativity also arises naturally in string theories with
background fluxes (see \cite{DN} and references therein).
Interestingly, gauge theories on
tori with magnetic flux and twisted Eguchi-Kawai models can be
reformulated in terms of non-commutative gauge theories\cite{GK}\cite{SZ}.

In the simplest example of a non-commutative space, the space-time coordinates satisfy
$$[x^i,x^j]=i\vartheta^{ij}$$
with $\vartheta$ independent of $x$. One of the most
striking differences with the commutative case  is the phenomenon of
ultraviolet-infrared mixing discovered in \cite{MvRS}: in non-commutative
spaces even massive theories can have amplitudes that are finite for generic
external momenta but diverge as these external momenta are taken to zero, a
divergence that one would usually classify as an IR phenomenon and that should
be absent in a massive theory if the Wilsonian renormalization group picture
holds.

UV/IR mixing has dramatic physical consequences. 
In the cases where UV/IR mixing
leads to pole-like singularities there are two kinds of possible behaviors
for the two-point function. The one-loop contributions can make the energies of
low momentum modes grow. This happens for example in non-commutative $\phi^4$ theory.
For these kind of theories it has been pointed out already in \cite{MvRS} that
a proper definition demands a resummation, not unsimilar to resummation techniques in
finite temperature field theories. For non-commutative $\phi^4$ this program has been carried
out explicitely in \cite{GP} where renormalizability of this model has been proven.

A rather different situation arises in non-commutative pure Yang-Mills theories \cite{Ha}, \cite{MST}. 
The one-loop correction
contributes with a negative sign and leads to modes with imaginary energies at at low momentum
\cite{LLT1}, \cite{RR}, \cite{LLT2},\cite{BGNV}. 
An important observation in this respect is that the non-planar one-loop graph can be
interpreted as the two-point function of the open Wilson-line operators of non-commutative
field theories \cite{KRSY1},  \cite{KRSY2},  \cite{AL},  \cite{KKRS}. Recently it has been argued that 
the instabilities
can be understood from a matrix model point of view as the non-cancellation of zero point
energies giving rise to a potential between the "D0"-branes of a non-supersymmetric matrix model
\cite{vR}.

In this paper we study similar phenomena on the
non-commutative torus \footnote{$^1$}{UV/IR mixing on the fuzzy sphere has been studied in \cite{CMS}.}. 
The UV-behavior of gauge theories on the non-commutative 
torus has been studied in \cite{KW} and a discussion of a scalar $\phi^3$ model and
gauge theory in the limit $\vartheta \rightarrow 0$ has appeared in \cite{GMW}. 
In the compact case $\vartheta$ which has
dimension length squared can be turned into a dimensionless parameter
$\theta$ after dividing by the volume of the torus. For rational $\theta$, field
theories on the non-commutative torus can be mapped by Morita equivalence to
theories of matrix valued fields on an ordinary, commutative torus with
twisted boundary conditions. For the
detailed construction, see for example \cite{GT} or \cite{S}. For a discussion
of rational/irrational $\theta$, see \cite{AB}\cite{H}.

The field theory on this commutative space is equivalent to the
non-commutative theory and yet should have properties that one would expect of
local field theories. We will discuss why UV/IR mixing does not introduce new
dependence on external momenta in the commutative theory with twisted boundary
conditions after integrals over loop momenta in the non-compact theory have
been replaced by discrete sums over modes appropriate for the torus. In this
analysis we will find that even the commutative theory displays some
unexpected behavior. For certain parameters, the renormalized dispersion
relation will have tachyonic modes, as it was the case for the theory on the
non-commutative plane.

The organization of the paper is as follows. In section two we discuss Morita
duality and introduce conventions and notation. In section three we compute
the one-loop corrections to the two-point functions in a scalar field model
with Moyal-bracket interactions in dimensions $D=2,3,4$. The results are
interpreted with respect to Morita duality and we also explicitely show how
the two-point function develops discontinuities in $\theta$ at one-loop. In
section four we discuss gauge theories. This turns out to be algebraically
much more involved than the scalar field, however we are able to show that
qualitatively the behavior of gauge theories is analogous to what happened in
the scalar case.  In section five we interpret the results of the previous
section with respect to UV/IR mixing and Morita duality. In section six we analyze the consequences
of the tachyonic mass terms we found for the phase structure of the gauge
theory in $D=4$ with two non-compact dimensions.  We argue that these
instabilities arise from the spontaneous breaking of translation invariance in
the non-commutative theory. In the Morita dual picture this corresponds to
condensation of electric fluxes via spontaneous breaking of $\MZ_N\times\MZ_N$
symmetry.  Some technical aspects of the calculations are collected in the
appendices.

\chapter{The non-commutative torus and Morita equivalence}
In this chapter, we will introduce Morita equivalence and explain our
conventions. We assume that two dimensions
(denoted by $x^1$ and $x^2$) are compactified on a non-commutative square torus
of radius $R$,
whereas the remaining $d=D-2$ are non-compact, commuting directions. For the
non-commuting directions, we have
$$[x^i,x^j] = i\vartheta^{ij}=2\pi i R^2 \theta\epsilon^{ij}.$$
Note the difference between $\vartheta$ which is a dimensionful quantity and
$\theta$ which is dimensionless. The distinction between rational
and irrational makes only sense for a dimensionless quantity. We are primarily
interested in the case of $\theta$ being rational.
Furthermore, here $\epsilon^{12}=-\epsilon^{21}=1$. This commutation relation
translates to a *-product
$$(f*g)(x) = \left.\exp\left(\pi i R^2\theta\epsilon_{ij}\partial^i_x
\partial^j_y\right)f(x)g(y)\right|_{y\to x}.$$
A field on the non-commutative torus can be
expanded as
$$\phi(\vec x)=\sum_{\vec k\in\MZ^2}\phi_{\vec k} e^{i{\vec x\over R}\cdot
\vec k}.$$
The *-product in terms of Fourier modes is then
$$(\phi*\psi)(x)=\sum_{\vec k\in\MZ^2}\left(\sum_{\vec l\in\MZ^2}\phi_{\vec
l}\,\,\psi_{\vec l-\vec k} e^{\pi i \theta \vec k\times\vec l} \right) e^{i{\vec
x\over R}\cdot\vec k}.$$
We can use the Baker-Campbell-Hausdorff formula
$e^Ae^B=e^{A+B}e^{\br 12[A,B]}$
to split the exponent of non-commutative coordinates into two factors. 
For ${\cal U}=e^{i{x_1\over R}}$ and ${\cal V}=e^{i{x_2\over
R}}$ we find the commutation relation
$$ {\cal U*V}={\cal V*U}e^{-2\pi i\theta}.\gln\UVrel$$
This implies
$${\cal U}^{k_1}*{\cal V}^{k_2} e^{\pi i \theta k_1k_2}=e^{i{\vec x\over R}\cdot \vec k}.$$
In the case of rational $\theta=p/N$ the commutation relation \UVrel\ can
also be realized in terms of $N\times N $ clock and shift matrices
$$U=\pmatrix{1&&&\cr
&e^{2\pi ip\over N}&&\cr
&&\ddots&\cr
&&&e^{2\pi ip(N-1)\over N}\cr},\qquad
V=\pmatrix{0&1&&\cr
&0&1&\cr
&&\ddots&\ddots\cr
1&&&\cr}.\gln\PQdef$$
They obey the same algebra as the non-commutative Fourier modes $\cal U$ and $\cal V$. Thus, instead of a
scalar field on the non-commutative torus we can equivalently study the matrix
valued field
$$\hat\phi(x) = \sum_{\vec k\in\MZ^2}\phi_{\vec k}U^{k_1}V^{k_2}e^{\pi i
\theta k_1k_2}e^{i{\vec x\over R}\cdot \vec k}\gln\matrixfield$$
on a commutative torus (with ordinary commuting coordinates $\vec x$).

The definition of the matrix generators \PQdef\ implies that
$U^N=V^N=\ID$.
The matrices $U^{k_1}V^{k_2}$ for
$k_1,k_2=0,\ldots,N-1$ are a basis of the $N\times N$ matrices. 
The mode
indices $\vec k\in\MZ^2$ thus label both matrix entries and momentum modes in
\matrixfield. It is therefore useful to decouple these two meanings. One should decompose the labels
as 
$$k_i = N \, r_i +\kappa_i\gln\splitmomenta$$
with $\vec r \in\MZ^2$ and $\vec\kappa\in\MZ_N^2$.
We will use Greek letters in the following for the ``fractional'' part of the momenta.
In terms of these \matrixfield\ becomes
$$\hat\phi (x) = \sum_{r\in\MZ^2}  \sum_{\kappa\in\MZ^2_N}
\tilde\phi_{\vec r,\vec\kappa} U^{\kappa_1}V^{\kappa_2}e^{\pi i
\theta \kappa_1\kappa_2}\exp\left({i{\vec
x\over R}\cdot \vec \kappa}\right)\exp\left({i{\vec x\cdot\vec r}\over R/N}\right),$$
where we changed the sign convention to $\tilde\phi_{\vec
r,\vec\kappa}=(-1)^{p(Nr_1r_2+r_1\kappa_2+r_2\kappa_1)} \phi_{N\vec
r+\vec\kappa}$. This should be read as follows: the last factor is the usual
Fourier mode for fields on an $R/N\times R/N$ torus with discrete momentum
$\vec r$. The sum over $\vec\kappa$ together with the matrix factor
$U^{\kappa_1}V^{\kappa_2}$ is the basis decomposition of $N\times N$
matrices. Note especially that
$\vec\kappa=(0,0)$ is the trace degree of freedom that we will also denote as
the $U(1)$ part of the matrix valued field. We conclude that only
the integer part $r_i$ of $k_i/N$ should really be thought of as momentum.  We
see that the commutative torus is smaller by a factor $N\times N$ than the
non-commutative one. This is not surprising since upon this rescaling also the
``density of degrees of freedom'' is kept constant as now we are dealing
with $N\times N$ matrices instead of scalars. The integral of the Lagrangian 
should be replaced by the integral of the trace of the new matrix valued Lagrangian.

However, $\hat\phi$ is not periodic when carried around the cycles of the
smaller torus, rather there are holonomies due to the phase $\exp({i{\vec x\over
R}\cdot \vec \kappa})$:
$$\eqalignno{\hat\phi(x_1+2\pi R/N,x_2)&=V^m\hat\phi(\vec x)V^{-m}\egln\boundaryone\cr
\hat\phi(x_1,x_1+2\pi R/N)&=U^{-m}\hat\phi(\vec x)U^m\egln\boundarytwo\cr}$$
where we defined $m={1\over N\theta}$. This of course should be interpreted
properly in $\MZ_N$. For $\theta=\br pN$ this means $pm\equiv 1(\hbox{mod}\,
N)$ or $pm-aN=1$ for some integer $a$ (this exists as $p$ and $N$ have no
common divisors).
It turns out\cite{vB1}\cite{S}\ that in gauge theories modes with fractional momentum
$\vec\kappa$ carry an electric flux
$$e_i\equiv p \epsilon_{ij}\kappa_j \,\mod N.\gln\elflux$$
This relation can be inverted as $\kappa_i\equiv m \epsilon_{ij}e_j \,\mod N$.

\ignore{
\chapter{The non-commutative torus and UV/IR-mixing}
In this section we will recall some well known facts about field theories on
the non-commutative torus. We will use coordinates $x_1$ and $x_2$ that are
periodic, that is 
$$x_i \sim x_i + R$$
(most of the time we will scale $R$ to unity) and that do not commute:
$$[x_i,x_j]=iR^2\theta\epsilon_{ij}.$$
Often, it is convenient to write this as
$$e^{i{x_1/ R}}\,e^{i{x_2/ R}} = e^{i{x_2/ R}}\,e^{i{x_1/
R}}\, e^{i\theta}$$
in terms of a Fourier basis. Any field $\phi$ (we will give explicit
expression for scalar fields but those of course will also hold for the
components of higher spin fields) can be decomposed in momentum modes as
$$\phi = \sum_{\vec k\in\MZ^2}\phi_{\vec k}\, e^{-i{\vec k\cdot\vec x\over
R}}.$$ 
In the case of rational $\theta=p/N$, one can use Morita equivalence to
translate this to a matrix valued field on an ordinary, commutative torus. To
this end, we use the $N\times N$ clock and shift matrices 
$$Q=\pmatrix{1&&&\cr
&e^{2\pi ip\over N}&&\cr
&&\ddots&\cr
&&&e^{2\pi i(N-1)\over N}\cr},\qquad
P=\pmatrix{0&1&&\cr
&0&1&\cr
&&\ddots&\ddots\cr
1&&&\cr}\gln\PQdef$$
with commutation relation 
$$PQ=QPe^{2\pi i\over N}\gln\commrel.$$
Furthermore, it is convenient to define $U=Q$ and $V=P^p$. These have the
commutation relation
$$ UV = VU e^{2\pi i \br pN}$$
of the Fourier generators above.
In addition, we need an integer that obeys $c\equiv p^{-1} (\hbox{mod } N)$.
Then, we can define a $N\times N$ matrix valued field
$$\hat\phi(x)=\sum_{\vec k\in\MZ}\phi_{\vec k} Q^{-ck_1}P^{k_2}\exp\left(
-i\pi\br cN k_1k_2\right)\exp\left(-i{\vec k\cdot\vec x\over NR}\right).$$
(CHECK NORMALIZATIONS!!!) The matrix product of fields of this type implements
the non-commutative $*$-product in the sense that
$$\widehat{\phi*\psi}=\hat\phi\hat\psi$$
as it was shown in \cite{GT}.
The matrix valued fields however obey twisted boundary conditions when carried
around the cycles of the torus:
$$\eqalign{\hat\phi(x_1+2\pi R,x_2) &= V^{-1}\hat\phi(x_1,x_2)V\cr
\hat\phi(x_1,x_2+2\pi R) &= U\hat\phi(x_1,x_2)U^{-1}\cr}\gln\bndcond$$
As also the integral is preserved,
$${1\over N^2}\int d^2x\phi = \int d^2x \tr\phi$$
any theory on the non-commutative torus with rational $\theta$ can be
translated into a theory of matrix valued fields living on an ordinary
commutative torus.   
Note that the torus of the commutative theory has periodicity $R/N$, thus the
torus is smaller by $N\times N$. On the other hand, there are the same number
of components in a $N\times N$ matrix. So, the number of component fields is
the same in both cases, in the commutative case the fields on the larger torus
are distributed in components of the matrix valued field. This transformation
of space-time localization into internal degrees of freedom will become
important later on.
}

This relation among field theories on different (non)-commutative spaces is
an example of Morita equivalence.
In general, Morita equivalence on the non-commutative torus relates a torus
with $\theta$ to one with
$$\theta' = {a\theta+b\over c\theta+d} \quad\hbox{for}\quad\pmatrix{a&b\cr
c&d\cr}\in SL(2,\MZ).$$
The radius, the size of the matrices and the twist transform as
$$R' = (c\theta+d) R\qquad \pmatrix{N'\cr m'\cr}=\pmatrix{a&b\cr
c&d\cr}\pmatrix{N\cr m}.$$
Rational $\theta$ can be related to $\theta'=0$, that is to a
theory on a commutative space. This is the relation we deal with in this note.

For concreteness,  we will first consider scalar field theories of the type
$${\cal L} = \br 12\partial_\mu\bar\phi\partial^\mu\phi +\br 12\mu^2\bar\phi\phi+ a 
\phi*\phi*\bar\phi*\bar\phi+b\phi*\bar\phi*\phi*\bar\phi.\gln\ablag$$
Our results do not rely on the fact that we are dealing with a scalar
field, in fact they should be generally applicable for field theories on the
non-commutative torus. Rather, we choose this example for notational
simplicity. In a later chapter, we also discuss
the case of pure gauge theory.
\goodbreak
At the one-loop level, the $a$ and $b$ type interactions receive different
contributions from planar and non-planar diagrams, as shown in Fig.\ 1.
However, most of the time, we will specialize to the Moyal-bracket interaction
$$[\phi,\bar\phi]^2_*$$
since this resembles the form of the gauge theory interaction.
This interaction term can be obtained from \ablag\ by stetting
$b=-a \equiv g^2/2$. From Fig.\ 2, we see, that for this choice the planar and non
non-planar graphs contribute with weights $2$ and $-2$. This is especially
convenient in the case of massless theories that might have (ordinary) IR
divergencies: because of the relative minus sign between the planar and
non-planar contribution these IR divergencies cancel and we can ignore
them. 
\epsfxsize=10cm
\psfig{The contributions of $\phi\phi\bar\phi\bar\phi$ and
$\phi\bar\phi\phi\bar\phi$ at one-loop order}{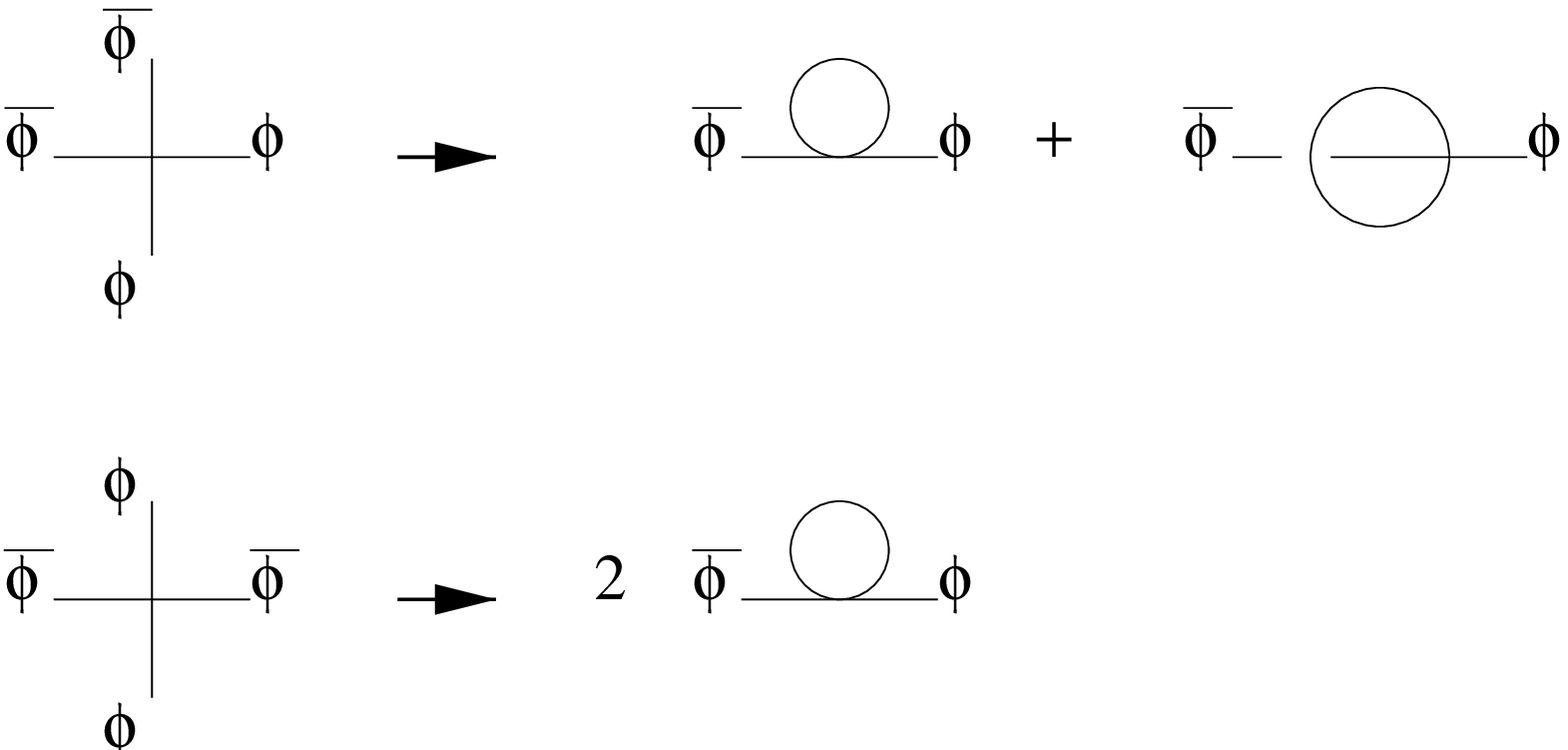}

It is well known\cite{GK}\cite{F} that the non-commutativity affects planar
diagrams only through phase factors for the external legs. In contrast,
non-planar diagrams contain phase factors that mix external momenta and loop
momenta, thus the effect of non-commutativity is much more severe.  In most of
what follows we will just give the expressions for the non-planar
amplitudes. The corresponding planar amplitudes can then be read of by setting
$\theta$ to zero in the Feynman integral. The total amplitude is just the
difference for the Moyal-bracket interaction. For more general choices of $a$
and $b$ in \ablag, one has to use the expressions shown in Fig.~1 to derive
the correct relative factors.

In
\cite{MvRS}, non-planar one loop diagrams were studied. The simplest example
being
\epsfxsize=3cm 
$$\vcenter{\hbox to 3cm{\epsfbox{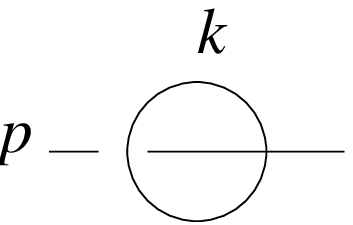}}}=\int {d^dk\over(2\pi)^d}
{e^{ip_\mu\theta^{\mu\nu}k_\nu}\over k^2+\mu^2}\gln\feynint$$ 
The exact value for this integral will of course depend on the dimension $d$,
but it is easy to qualitatively anticipate the result: without
non-commutativity, the mass makes the integral well behaved for small $k$. For
large $k$ the integral will diverge logarithmically for $d=2$ and like
$k^{d-2}$ for higher dimensions.

In the presence of the phase factor, the integrand oscillates and in fact the
integral converges for large $k$. The combination $p\circ p :=
p_\mu\theta^{\mu\nu}\theta_{\nu\rho}p^\rho$ has dimension length squared. As a
dimensionful quantity, it can serve as a regulator. In fact, at leading order,
the two dimensional integral behaves like $\log (p\circ p)$ and in higher
dimensions like 
$$1\over (p\circ p)^{d-2\over 2}.$$
We see, that the non-commutativity introduces a non-trivial dependence on the
external momentum. Of course, if there is no external momentum in the
non-commutative directions, that is $\theta^{\mu\nu}p_\nu=0$, this regulator
vanishes and the integral again diverges.

This non-trivial dependence and especially the divergence for $p\to 0$ for
massive theories came as a surprise in \cite{MvRS} and has been termed
``UV/IR-mixing''. The ultraviolet divergencies reentering for infrared values
of external momenta even casts doubt on the validity of the Wilsonian
renormalization group in non-commutative theories.

On the other hand, as we have explained above, on the torus with rational
values of $\theta$, the non-commutative theory can be reformulated in terms of
a commutative theory with matrix valued fields and twisted boundary conditions. This
commutative theory is not expected to contain any such surprises and thus
should not show any signs of UV/IR-mixing. It is one of the topics of this
paper to resolve this puzzle.

\chapter{One loop on the torus}
Now, that we have set the scene, we can perform the actual calculation in the
scalar field model in $2+d$ Euclidean dimensions.
The interesting non-planar amplitude \feynint\
takes then the form
$$\eqalignno{
A(\vec{n})& = \int {d^dk\over (2\pi)^d}{1\over (2 \pi R)^2} \sum_{\vec l\in \MZ^2} 
{ e^{i 2\pi \theta l_j n_k \epsilon_{jk}}
\over k^2 + {\vec{l}^2\over  R^2}+\mu^2 }\cr& =  \int {d^dk\over (2\pi)^d}{1\over (2
\pi R)^2} \sum_{\vec l\in \MZ^2}
\int_0^\infty d\alpha e^{ -\alpha(k^2+{\vec{l}^2\over R^2}+\mu^2) +i 2\pi  \theta l_j n_k 
\epsilon_{jk}} \,.\egln\amptorus}
$$
The inflowing momentum is $\vec{n}/R$. Note that the amplitude does
not depend on the external momentum in the commutative directions.  Here, we
have introduced a Schwinger parameter $\alpha$. In the following we will use $b_j = \theta \epsilon_{jk} n_k$

Now we can integrate out the momenta corresponding to the commuting directions. 
After a Poisson resummation the amplitude can be written as
$$
A = {\pi^{d/2+1} \over (2\pi)^{d+2}} \int_0^\infty d\alpha \alpha^{-1-{d\over 2}}
\sum_{\vec l\in \MZ^2} e^{ - {(\vec{l}+\vec{b})^2 R^2 \pi^2\over
\alpha}-\alpha \mu^2}\,\gln\ampmass.
$$ 
Let us first treat the massless case. We finally integrate over the Schwinger parameter and arrive at
$$
A = {\pi^{1-{d\over 2}}\over (2\pi R)^{d}} { \Gamma({d\over 2})\over (2\pi)^2} \zeta(d/2,\vec{b})\,.
\gln\ampzeta$$
In the last line we introduced the Epstein zeta function
$$
\zeta(s,\vec{b}) = \sum_{l\in \MZ^2}^{} {}^{'} {1\over [(\vec{l}+\vec{b})^2]^s}\,.
\gln\zetasum$$
We see that the only effect of the non-commutativity is to shift the momentum
lattice for the non-planar graph. All oscillating phase factors have been
removed by the resummation.

Of course this representation of the zeta-function as an infinite sum is only convergent
for $Re(s)> 1$. Moreover, the sum is defined in
such a way as to exclude lattice points where the denominator under the sum could vanish, 
i.e. for $\vec{b}=0$ we exclude the origin $\vec{l}=(0,0)$. This is denoted by the symbol $'$
in the sum.
It turns out that for $d>0$ this
is nothing but a UV-regularization since after the Poisson resummation the UV is $\vec{l}=(0,0)$.
We see that this UV-divergence is regulated by the non-commutativity. 
The situation at $d=0$ is slightly different as we will explain in the next paragraph.
We also
consider the vector $\vec{b}$ to be reduced on the lattice, $b_j \in (-{1\over 2},{1\over 2}]$.
This reduction on the lattice is evident from the definition of the amplitude in formula \amptorus\
since the integer part of the vector $\vec{b}$ does not contribute
to the expression. Alternatively, one can apply a shift to the sum over the dual
lattice in \zetasum.

As is well-known, the Epstein zeta function can be analytically continued over the complex 
$s$-plane with the help of Jacobi's theta-function in the form
$$\eqalignno{
\zeta(s,\vec{b}) =& {\pi^s\over \Gamma(s)}\left\{ {1\over s-1} - {\delta_{\vec{b},0} \over s}+\right. \cr
&+ \left.\int_1^\infty\!\!\!\! dt \left[ t^{-s} \left( \prod_{j=1}^2 \vartheta(it,b_j) -1 \right) +
\left(t^{s-1} e^{-\pi t \vec{b}^2}  \prod_{j=1}^2 \vartheta(it,i t b_j) - \delta_{\vec{b},0}
\right) \right] \right\}\,.\egln\defzeta}
$$
This representation is explained in appendix~A.

Taking together planar and non-planar graphs in the theory with Moyal bracket interactions,
the one-loop effective action can now be written as 
$$\Gamma_2^{(1)} =  k^2+{\vec n^2\over R^2}+{g^2R^{2-d}\Gamma({\br d2})\over
2^d\pi^{3d+2\over 2}R^2} 
\left(\zeta(d/2,\vec{0})-\zeta(d/2,\vec{b})\right)\,.
$$
Here $k^2$ denotes the momentum in the uncompactified dimensions (if
existent). If there is a non-compact direction we can undo the Wick rotation
and interpret the expression $\Gamma_2^{(1)}=0$ as a dispersion
relation. $g^2R^{2-d}$ is a dimensionless parameter that should be small in
order for perturbation theory to hold.

Whenever $\vec{b}=0$ planar and non-planar contribution cancel each other
exactly. This can be easily understood in terms of the Morita-dual theory.
Recalling the definition of $\vec{b}$
one sees that  $\vec{b}=0$ corresponds to momentum modes $\vec n=N\vec r$.
Morita duality maps these modes into
the momentum $\vec r$ modes of the overall $U(1)$ degree of freedom.
The $U(1)$ modes are non-interacting for Moyal bracket interaction and
therefore the one-loop correction vanishes.

The modes for which $\vec b\ne 0$ are mapped to the $SU(N)$ degree of freedom under Morita
duality. Since 
the one-loop correction depends only on the parameter $\vec{b}$
the amplitude is thus periodic under
$\vec n \rightarrow \vec n+N\vec r$, with $\vec r\in \MZ^2$.
Therefore, in terms of the Morita dual theory these corrections are mass
terms. 

Let us discuss now the poles in in the zeta function at $s=0$ and $s=1$.
In $D=2$, that is 
$s=0$ the zeta function vanishes except at the point $\vec{b}=0$. 
However, the gamma function in the denominator in \defzeta\ is canceled by the
Gamma function in the Amplitude in \ampzeta. Therefore the amplitude has a pole
for $\vec{b}=0$. The nature of this divergence is easily understood
by examining \amptorus. For $D=2$ this expression has a divergence arising
from the zero-mode on the torus ($\vec{n}=0$). The sum in \amptorus\ should then be defined 
by omitting the zero mode. It is this divergence that is regulated by the zeta-function
expression for the amplitude. At $D=2$ the zeta function regulates the IR but
leaves us with the UV divergence. 
We can now regulate the UV by subtracting the pole at $s=0$ in the brackets in \defzeta\
for the definition of the amplitude. \footnote{${}^2$}{Of course, this leaves us with a finite 
undetermined constant. This is nothing but the usual mass
renormalization. Notice that this is a universal mass for all $SU(N)$ modes in
contrast to the corrections arising from the non-planar amplitude. }
\ignore{More precisely,
since we started with a massless theory at tree level a mass term is generated by the
one loop correction. The planar amplitude should then be defined in the same way.}

There is no pole for $D=3$.
The situation with two commuting, compact dimensions ($D=4$) presents a new feature. 
There is a pole in \defzeta\ at $s=1$. This can be understood as follows. 
At low energies the
theory reduces effectively to a two-dimensional field theory. Thus this field theory will have an
infrared divergence as is common in two dimensions. We will not further worry about
this infrared divergence since the form of the Moyal-bracket interaction provides automatically 
an IR-regularization. Indeed the poles at $s=1$ cancel between the planar and non-planar
contributions. Basically this is the statement that the theory is non-interacting at distances
large compared to the radius of the torus. There is however a subtlety here. Due to one-loop effects 
the modes that are important at low energies are not only the zero modes on the torus. Again the contribution
of the non-planar diagram can lower the energy of a momentum mode on the torus significantly.
So there might be finitely many other modes that are important in the infrared
as we will now discuss. 

It can happen that for low momentum modes and large $N$ (large gauge group in the Morita dual 
theory) the contribution of the non-planar modes does not only dominate the planar one but also
becomes larger than the tree-level term. If $\theta \epsilon_{jk} n_k$ is very
close to a lattice point $\vec{b}$ becomes very small and for $\vec{b}\rightarrow \vec{0}$
the $\zeta$ function grows without bound. Expanding the effective action for
small $\vec b$ we find
$$
\eqalignno{\Gamma_2 \approx& {\vec{m}^2\over R^2} + \frac{2g^2}{\pi} \log( |\vec{b}| ),\qquad D=2 \cr
\omega^2 \approx & {\vec{m}^2\over R^2} - {g^2\over 2 \pi^2 R} {1\over |\vec{b}|},\qquad D=3\cr
\omega^2 \approx & k^2 + {\vec{m}^2\over R^2} - {g^2\over 4 \pi^4 R^2}
{1\over \vec{b}^2},\qquad D=4.
 }
$$
Since the contribution of the non-planar diagram
has negative sign this means that mass squared becomes negative.

It is convenient to label the modes by $\vec e$ according to \elflux\ even if
the interpretation as ``electric'' flux only makes sense in the gauge theory.
We see then that $\vec b=\br 1N \vec e$.  Therefore the non-planar
contribution becomes maximal for unit electric flux in the Morita dual
language. This does not necessarily mean that the unit electric fluxes are
also the ones that become unstable. It might happen that $e_i=1$ corresponds
to a rather large $n_i$. The tree level term might then still be larger than
the one-loop correction resulting in a positive inverse two-point function for
this mode. In this case unstable modes can first appear for larger electric
fluxes.

\epsfysize=2.3 true in
\psfig{This graph shows the dispersion relation for $\theta={1000\over 3001}$.}{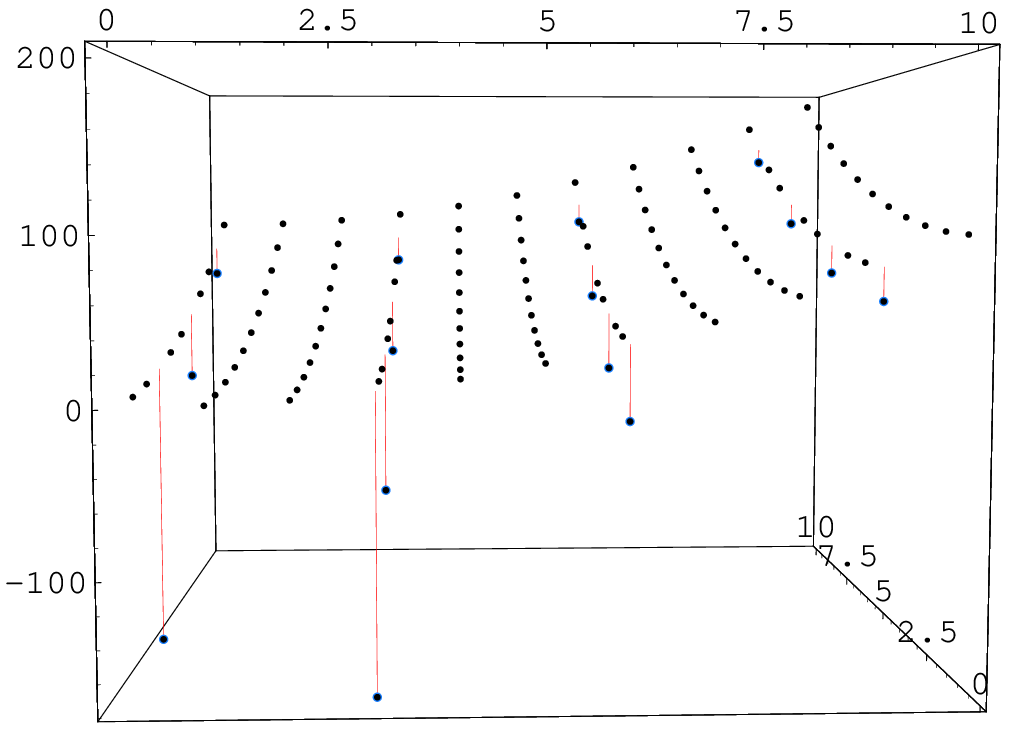}

To illustrate this point let us work out a concrete example in $D=4$.
The graph in Fig. 2 shows a typical plot of the dispersion relation. We chose $\theta=1000/3001$ such that
$\vec{b}$ becomes very small for every third mode in the momentum lattice on the non-commutative torus.
For low lying modes the one-loop correction dominates the tree level term and the mode becomes unstable.
The (non-commutative) momentum modes corresponding to unit electric flux are in this case $n_i=3$.
If we keep the radius $R$ and the coupling constant $g$ fixed and take $\theta=1000/3003$ instead the 
unit electric flux is mapped to the momentum mode $n_i = 1000$.
Although this is the mode for which the one-loop correction is maximal it is not necessarily tachyonic 
since the tree-level term is much larger. The most tachyonic modes will still be the momentum modes
$(3,0)$ and $(0,3)$ as before (corresponding to the electric flux $e=3$). 
Of course had we chosen $\theta=1/3$ instead, the corrections for every third mode would vanish identically.
\ignore{This discontinuity in $\theta$ is further shown in Fig.3 where the squares of the energies for the
modes $(3,n)$ with $n\in\{0,1,2\}$ are plotted. The mode $(3,0)$ receives large correction for
$\theta$-values near $n/3$ whereas the other modes behave rather smoothly except at $\theta=0$ and
$\theta=1$. Both values correspond to the limit of a non-interacting theory with $U(1)$ symmetry.
At $\theta=0$ this is obvious whereas at $\theta=1$ it is the Morita-dual that leads to
a free $U(1)$ theory. 
\epsfxsize=6.5in 
\psfig{Discontinuity of $\omega^2$ as a function of $\theta$.}{thetadiscontinuity.ps}
\bigskip
}

Let us compare our findings with the behavior in the non-compact case where it
is well known that low momentum modes can become tachyonic at
one-loop\cite{LLT2}\cite{RR}\cite{BGNV}.  We could have expected that at least
for very large radius the theory on the torus should behave similar as the
non-compact one.  Of course in the non-compact case there is no mass-gap and
the one-loop contribution grows arbitrarily large for low momentum. On the
torus the zero mode behaves completely benign. Nevertheless low, but non-zero,
momentum modes can become unstable. For large
momenta the tree level term grows quadratically and will certainly dominate
over the one-loop correction. What is perhaps somewhat surprising here is the
fact that tachyonic behavior appears even in the case with rational
$\theta$. One might have expected that this peculiar feature of
non-commutative theories is absent here since there is a Morita dual
description in terms of a conventional $U(N)$ theory on a commutative torus,
where one does not naively expect any UV/IR mixing. We emphasize that 
instabilities appear only for large (but finite) $N$.

In the case of irrational $\theta$, integer multiples of $\theta$ get
arbitrarily close to integers. The question of course remains whether the
one-loop correction can compete with the tree level term that is quadratic in
the external momenta. The ``pigeon hole theorem'' of rational approximation
theory implies that there are infinitely many integers $n$ such that $n\theta$
is closer to an integer than $1/n^2$. That is, in $D>4$ there are always
infinitely many unstable modes. For $D=4$, the two terms are generically of
the same order and the existence of unstable modes irrespective of $\theta$ 
depends on the coupling $g$. For $D<4$ and for generic $\theta$ there need not
be unstable modes. However, there are irrational numbers $\theta$ such that
the estimate $1/n^2$ above can improved to any power $1/n^a$ and hence for
those values of $\theta$ there are infinitely many unstable modes in all
dimensions.

Let us now have a look at the massive case.
Integrating the amplitude \ampmass\ over $\alpha$ gives
$$
A = {(R \mu)^{d\over 2} \over (2\pi) (2 \pi R)^d } \sum_{\vec{l}\in \MZ_2}^{} {}^{'} 
{1\over [(\vec{l}+\vec{b})^2]^{d\over 4} } K_{d\over 2} \left(2\pi \mu R \sqrt{(\vec{l}+\vec{b})^2}\right)\,.
$$
$ K_{d\over 2}$ denotes a modified Bessel-function. 
Since the Bessel-function decays 
exponentially for large $\vec l$,  there is no $1\over s-1$ pole
connected to an IR singularity.  The behavior at small
$\vec b$ is the same as in the massless case. 
In each of the cases for small enough $|\vec{b}|$ the one-loop correction dominates the tree-level term.
Independently of the presence of a mass-term we find therefore tachyonic instabilities.  

We have limited ourselves to the one-loop approximation. 
One must ask the question whether this is reliable. 
Notice that $\vec{b}$ acted as a UV-regulator. In the case $D=4$ we are
dealing with a renormalizable field theory. Therefore $n$-loop corrections to the two point function
will at most behave quadratically in the cutoff and therefore contribute terms
of the order $g^{2n} \over \vec{b}^2$. This is clearly sub-leading at small coupling. For $D<4$ the models
are super-renormalizable and therefore higher loop terms will scale with lower powers of the cutoff than the
one-loop term. In the non-compact case there is however an additional source for divergencies. It stems from 
the fact that insertions of the non-planar one-loop graph into higher-order loop graphs can lead to new
divergencies due to the infrared singular behavior of the one-loop correction. From our results it is
evident that no such difficulty arises in the compact case with rational
$\theta$. 
Indeed the non-planar
one-loop contribution is perfectly non-singular on the complete lattice of
non-commutative momenta.
However, there could still appear terms of order $g^{2n} \over \vec{b}^{2n}$
in these graphs. As argued in \cite{MvRS} such difficulties can be handled by
resumming the propagator. In the concrete model we considered here the
resummed propagator leads to tachyonic pole. In this case one has to find
first the effective potential and its minimum before
proceeding in the loop expansion. In any case the tachyonic behavior
can not be ignored.

\chapter{Gauge Theory on the NC Torus}
In this section we will consider a $U(1)$ gauge theory on the 
non-commutative torus. If the non-commutative parameter
$\theta$ is rational, Morita duality relates this theory to
an ordinary $U(N)$ theory on a torus with magnetic
flux $m$. It is thus of special interest to study this case.

In order to analyze the quantum corrected dispersion relations
we need to evaluate the polarization tensor. Its non-planar part 
is given by
$$
\Pi_{\mu \nu}^{NP} = {g^2 \over (2 \pi R)^2} 
\sum_{\vec{l} \in\MZ^2} \int
{d^d k \over (2 \pi)^d} \; {{\rm cos} (2 \pi \theta \, l_i n_j \epsilon_{ij})
\over 
K^2 \, (K-P)^2}\;
f_{\mu \nu} (P_\rho,K_\sigma) \, , \gln\polfirst
$$
where as before $d$ is the number of commutative non-compact directions.
We denote $\mu=(\alpha,i)$, with $\alpha=1,..,d$ labeling the 
non-compact dimensions and $i=1,2$ the two directions of the
non-commutative torus. 
The inflowing momentum is $P_\mu=(p_\alpha,
n_i/R)$ and the loop momentum $K_\mu=(k_\alpha, l_i/R)$.
In this section capital letters denote $D$ dimensional momenta.
From now on we will set $R=1$ in order to simplify our expressions;
it can be reintroduced by dimensional analysis. 
The polynomial $f_{\mu \nu}$ is \cite{PS}
$$f_{\mu\nu} =
4d \, K_\mu K_\nu - 2d \, (P_\mu K_\nu + P_\nu K_\mu) 
+(d-4)\, P_\mu P_\nu 
- g_{\mu \nu}\, \big( 2d \,K^2 -(4d +2)\, K\! \cdot \!P+(2d-3)\,P^2 \big) \, .
$$
Integrating the loop momentum along the non-compact directions,
Poisson resumming and using standard properties of Bessel functions,  
we can rewrite \polfirst\ as
$$
\Pi_{\mu \nu}^{NP} =  {g^2 \over (4 \pi)^{{d \over 2}+1}}  
\sum_{\vec{l} \in\MZ^2} \, 
\int_0^1 \!d x \, {\rm cos}(2 \pi P\cdot L x) \! \int_0^{\infty} {d \alpha \over \alpha^{d \over 2}}\,
 e^{- \alpha x(1-x)P^2 -
{\pi^2 \over \alpha} L^2}  
h_{\mu \nu}(P_\rho,L_\sigma) \, , \;\; \gln\pol
$$
where we have introduced $L_\mu=\big( 0,l_i- \theta \epsilon_{ij}n_j \big)$ and
$$\eqalignno{
h_{\mu \nu}(P_\rho,L_\sigma) =&  \;- {4d \,\pi^2 \over \alpha^2} 
\left[  L_\mu L_\nu 
\, - \, \big( P_\mu L_\nu + P_\nu L_\mu - g_{\mu \nu} P\! \cdot\! L \big)
\, {P \! \cdot\! L \over P^2} \, \right] \, 
-  \cr
& - \big( \, P_\mu P_\nu - g_{\mu \nu} P^2 \big) \, 
\big(\, 4d \,x\, (1-x)-d+4 \, \big)  \, . \egln\seriesb}
$$
We refer to Appendix~B for a detailed derivation of these expressions.
The planar part of the polarization tensor can be obtained from \pol-\seriesb\ 
by setting $\theta=0$.
Notice that both terms in the rhs of \seriesb\ are manifestly transverse,
so that $P^\mu h_{\mu \nu}=0$. 
The second term is proportional to the tree level two-point function.
It contributes to the effective coupling constant and
does not modify the form of the dispersion relation at 1-loop order. 
We will concentrate thus in the first term. Notice that this term
is proportional to the number of degrees of freedom of the gauge field,
i.e. $d=D-2$. Thus the cases of interest in this section will be 
$d=1,2$.

The $\alpha$ integration in \pol\ can be easily performed
$$\eqalignno{
\int_0^{\infty} {d \alpha \over \alpha^{{d \over 2}+2}} & \, 
e^{- \alpha P^2 x(1-x)-{\pi^2 \over \alpha} L^2} \,= \cr 
& = \, 2 \left( {\sqrt{P^2 x(1-x)} \over \pi |L|} 
\right)^{{d \over 2} +1}
K_{{d \over 2}+1} \big(2 \pi |L|  
\sqrt{P^2 x(1-x)} \big) \, .
\egln\bessel }$$
We will approximate the sum over $\vec{l}\in \MZ^2$ by 
keeping only the vector $\vec{l}_0$ at which $|L|$ acquires its 
minimum. This approximation is best for external
momentum $\vec{n}$ such that $|L| \sim 1/N$. In analogy with the 
previous section we denote the value of $L$ at $\vec{l}_0$ by 
$B_\mu=(0,b_i)$, where 
$b_i=(l_0)_i-\theta \epsilon_{ij}n_j $. These are the momenta for 
which we expect large 1-loop corrections. 
Furthermore we assume that $|B||P|$ is small. It turns out that the solution to the 1-loop 
corrected dispersion relation satisfies this requirement. 
The expansion of the modified Bessel function \bessel\ at small values 
of the argument is
$$
K_{{d \over 2}+1}(2z)={\Gamma({d \over 2}+1) \over 2 z^{{d\over 2}+1}}-
{\Gamma({d \over 2}) \over 2 z^{{d \over 2}-1}} + \ldots \, . \gln\expr
$$
Given that \seriesb\ is transversal, each
contribution to the polarization tensor derived from expanding
$K_{{d \over 2}+1}$ will satisfy independently the Ward identity.

Denote the contributions of
the polarization tensor associated with the two terms in \expr\  
by $\Pi^{NP,1}$ and $\Pi^{NP,2}$. Substituting
the expansion into \pol, we obtain
$$
\Pi^{NP,1}_{i j} =  {d \,  g^2 \, \Gamma\big( {d \over 2}+1 \big) \over 2^d \, 
\pi^{{3 \over 2}d+1}}\,  {b_i b_j \over  \;\;\;
|\vec{b}|^{d +2}} \; \delta_{0,\vec{n}\cdot \vec{b}} \;\;\; ,
\gln\main
$$
with all other components vanishing. The delta function arises from the $x$-integration of 
${\rm cos}(2 \pi P\! \cdot \!B\, x)$, since $P\cdot B=\vec{n}\cdot \vec{b}$. 
The non-compact limit is achieved by setting $\theta=1/N$ with 
$N\rightarrow \infty$. In this limit it is immediate to see that \main\
reproduces the polarization tensor on $\MR^d \times \MR_{\cal \theta}^2$
\cite{Ha},\cite{MST}.

From the next to leading term in \expr\ we obtain
$$
\Pi^{NP,2}_{\mu \nu} = {d \,  g^2 \, \Gamma\big( {d \over 2} \big) 
\over 2^d \pi^{{3 \over 2}d-1}}\; { c(\vec{n} \cdot \vec{b})  \over
|\vec{b}|^d} 
 \, \left[ B_{\mu} B_{\nu}  \, P^2 
 -  (P_\mu B_{\nu} + P_\nu B_{\mu} - g_{\mu \nu} P\! \cdot \!
B ) \, P\! \cdot \! B\right]  \,  , \gln\sub
$$
where 
$$
c(\vec{n} \cdot \vec{b})=\int_0^1 dx  \, x(1-x)\, {\rm cos} (2 \pi x \vec{n} \cdot
\vec{b}) = \, \left\{ 
\!\! \matrix{
& {1 \over  6} \;\;\;\;\; & \vec{n} \cdot \vec{b}=0 \cr  
 & -{1 \over 2 \pi^2 (\vec{n} \cdot \vec{b})^2} &  \vec{n} \cdot \vec{b}\neq 0 } \right.
$$
The contribution at $\vec{n} \cdot \vec{b}=0$ has the same structure as \main\
but is clearly sub-leading with respect to it. For $\vec{n} \cdot \vec{b} \neq
0$ \sub\ shows a more involved structure, whose origin is the
requirement of preserving transversality. These modes 
also suffer corrections of order $g^2/|\vec{b}|^d$.
Thus we see that regardless whether $\vec{n} \cdot \vec{b}$ vanishes or not
there are always corrections of this order. Further terms in the 
expansion of the Bessel function at small argument are  
sub-leading.

We will study now the dispersion relations. The
quantum corrected inverse propagator is given by 
$$ 
\Gamma^{(1)}_{\mu \nu}=(\omega^2-{\vec n}^2)g_{\mu \nu} - 
\Pi^{NP,1}_{\mu \nu}-
\Pi^{NP,2}_{\mu \nu} \, . \gln\drt
$$ 
We have assumed that the momentum along the spatial 
non-compact direction vanishes.
Finding the poles of the propagator is involved 
because of the complicated, non-diagonal structure of
$\Pi^{NP,2}$. We will not attempt to solve the
problem in the general case. Our aim is to analyze the 
possible appearance of unstable modes. 
This can only happen if the quantum corrections dominate over the 
tree level term. We will see that this condition is never met
for modes with $\vec{n} \cdot \vec{b} \neq 0$. When $\vec{n} \cdot \vec{b}
\neq 0$ the following relations hold
$$
|\vec{b}|=|\vec{l}_0-\theta \epsilon\!\cdot \! \vec{n} | > 
{|{\vec l}_0 \cdot {\vec n}| \over |\vec{n}|} \geq {1 \over |\vec{n}|} \, . 
\gln\pain
$$
The  dominance of the 1-loop correction in \drt\ requires
$g^2/|\vec{b}|^d \geq \vec{n}^2$. 
For $d=1,2$ and small coupling, this is not possible for modes 
satisfying \pain. Thus we will concentrate in momentum
modes with $\vec{n} \cdot \vec{b} = 0$. In this case, $\Pi^{NP,2}$ is 
negligible with respect to $\Pi^{NP,1}$. The dispersion relation of the 
tranverse photon polarized along the non-commutative torus for these
modes reduces to
$$ 
\omega^2= {\vec{n}^2 \over R^2} - {d \,  g^2 \, 
\Gamma\big( {d \over 2}+1 \big) \over (2 R)^d \, 
\pi^{{3 \over 2}d+1}}\,  {1 \over |\vec{b}|^d} \, ,
\gln\light
$$ 
where we have reintroduced the dependence on the radius of the torus.
This relation is analogous to the one obtained for the 
scalar model and thus the analysis performed there applies. For sufficiently 
large $N$ there are momentum modes that acquire negative energy square, giving rise to
perturbative instabilities. Among the modes that satisfy $\vec{n} \cdot
\vec{b} = 0$ are those with momentum proportional to $(1,0)$ or $(0,1)$. 
It is easy to see that they are actually the first candidates to become 
perturbatively unstable. The 
analysis of the instability and its interpretation in terms of the
Morita dual theory are left for section 6.

\chapter{UV/IR mixing on the torus?}
One of the major surprises of field theories on the non-commutative
plane is the phenomenon of UV/IR mixing\cite{MvRS}. Loop amplitudes in these 
theories show an unusual dependence on momenta. Even massive 
theories that are normally well behaved in the infrared have
new divergences as $p\to 0$. 

Here we will discuss UV/IR mixing in view of our results on the
non-commutative torus for rational $\theta$. 
In this case there is a Morita equivalent commutative theory in terms
of matrix valued fields with twisted boundary conditions. As the two theories
are equivalent, one would  expect some manifestation of UV/IR mixing in terms
of the commutative theory. 
We are going to argue that
in fact on the torus the usual low energy analysis holds. 
Recall the splitting of the non-commutative momentum
as in \splitmomenta. The amplitudes in the commutative theory with twisted
boundary conditions do not show any unusual momentum dependence.
Especially there are no new divergencies in the low energy regime.
What appears as an unusual momentum dependence of loop amplitudes in the
non-commutative picture becomes a certain dependence on the electric fluxes,
or equivalently Lie algebra degrees of freedom, in the commutative description. 

The main difference between the non-compact and compact case is that
momentum integrations are replaced by discrete sums over the momentum
modes. This allowed us to apply the Poisson resummation to translate the
oscillatory factor to a shift in sum over modes. For concreteness let us
discuss the case with two compact and two non-compact directions. There we
derived for the one loop self-energy correction the expression \ampzeta. The
external momentum entered as 
$$\sum_{\vec k\in\MZ^2}{}^{'}{1\over(\vec
k-\theta\vec p)^2}$$
If we rewrite this for rational $\theta=p/N$ using the splitting
\splitmomenta, we see that possibly after a relabeling 
the sum only depends on the ``matrix entry'' part $\pi_i$ of the momentum
$\vec p = N \vec r+\vec\pi$:
$$\sum_{\vec k\in\MZ^2}{}^{'}{1\over(\vec k-\br pN\vec \pi)^2}$$
This rewriting demonstrates explictly that there is no unusual dependence on
the commutative torus momenta $\vec r$. Thus this amplitude
does not show UV/IR-mixing.

Of course not all amplitudes depend only on the fractional momenta. 
This is only the case if the corresponding planar amplitude is independent of
the external momenta. 
Generically, 
dependence on momenta and electric fluxes mixes.
This happens for example in the gauge theory two point function 
or for the one loop correction to
the four point function (the fish-graph). Let us consider this case
in a bit more detail. 
\epsfxsize=4cm
\psfig{The fish graph}{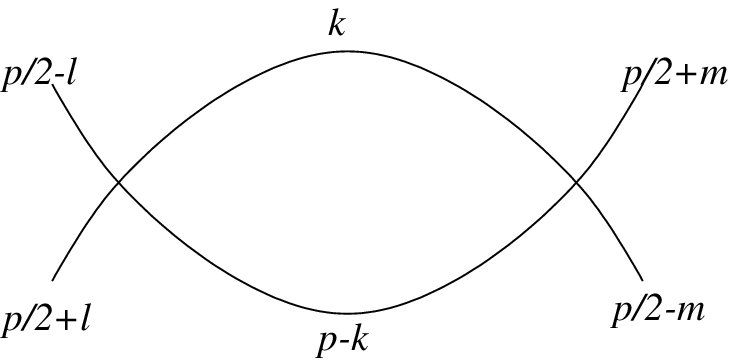}

The amplitude after Poisson resummation is proportional to 
$$
\sum_{\vec k\in\MZ^2}\int_0^1dx\int_0^\infty d\alpha
e^{-{\pi^2\over\alpha}(\vec k-\theta\vec p)^2+2\pi i (1-x)\vec p\cdot(\vec
k-\theta\vec p) +\alpha(x^2-x)\vec p^2}.\gln\fishamp$$
The external momentum $\vec p$ appears in two different ways: by itself and in
the combination $\vec k-\theta\vec p$. The latter reduces to $\vec k$ in the
commutative case $\theta=0$ so the former is the usual momentum dependence of
this amplitude in a commutative theory. The only effect of the
non-commutativity is in the combination with $\vec k$.

Again, for rational $\theta$ and after splitting $\vec p$ into its momentum
part $\vec r$ and its matrix entry part $\vec \pi$ the effect of $\vec r$ in
these terms can be completely undone by an integer shift in the summation
variable $\vec k$ thus it does not influence the UV behavior (that is $\vec
k=0$ after Poisson resummation) of the amplitude. Only $\vec \pi$, the
specification of the matrix entry, has an influence. This is of course
expected since the background flux that induces the twisted boundary
conditions breaks the degeneracy among the matrix entries. We see that from
the point of view if the commutative theory there is no unexpected momentum
dependence and thus no UV/IR mixing in the sense explained above.

We find that the regularization provided by the non-commutativity of
the coordinates when re-expressed in terms of the commutative theory depends
only on the specification of the matrix entry, and not on the
commutative momenta. Thus there are no new IR divergencies and
therefore no UV/IR mixing. 

\ignore{
By ``unexpected'' we mean the following: compare to the expression for
\fishamp\ in the case of vanishing $\theta$:
$$\sum_{\vec k\in\MZ^2}\int_0^1dx\int_0^\infty d\alpha
e^{-{\pi^2\over\alpha}\vec k^2+2\pi i (1-x)\vec p\cdot\vec
k+\alpha(x^2-x)\vec p^2}$$
Of course, this amplitude is not independent under shifts $\vec p\mapsto\vec
p+ N\vec l$ for an integer vector $\vec l\in\MZ^2$. The point is that
\fishamp, after a shift in the sum, has the same functional dependence under
such shifts as the expression for $\theta=0$. Thus \fishamp\ has the same functional
dependence on momenta as the theory without twisted boundary conditions and
therefore there is no UV/IR-mixing in the sense above of introducing a new
functional dependence on momenta.
We expect this
behavior to be generic and not restricted to the cases we have analyzed in
detail. 
}

\chapter{Spontaneously broken translation invariance and  electric flux
condensation.}
We have seen that perturbative tachyonic behavior arises at one
loop in the pure $U(1)$ gauge theory on a rational non-commutative
two-torus. This instability does not occur for the zero modes,  which
are dual to the free decoupled $U(1)$ modes of the Morita
equivalent $U(N)$ theory. 
The instability of the modes with
non-zero momentum suggests they get a non-zero expectation value
which spontaneously breaks translation invariance.  We shall see
that this occurs only for $D=4$,  whereas for $D<4$ the apparent
symmetry breaking is washed out by quantum fluctuations.
In this section we will discuss the phase structure of the theory
and its interpretation in the Morita equivalent $SU(N)/\MZ_N$
description\footnote{${}^3$}{Strictly speaking,  the Morita dual is a $U(N)$ theory,
however we generally ignore the trivial dynamics of the decoupled $U(1)$
sector.}.  The spontaneous breaking of translation invariance in the
non-commutative theory takes a much
more familiar form in the Morita equivalent
$SU(N)/\MZ_N$ description, namely that of electric flux condensation.
A closely related phenomenon arises in finite temperature
deconfinement,
which is indicated by the spontaneous breaking of a $\MZ_N$ symmetry of
large gauge transformations on a compactified Euclidean ``time''
direction \cite{SY1}\cite{SY2}.

The properties of the map between the non-commutative theory and the
$SU(N)/\MZ_N$ theory which we shall require (see section 2) are the following.
For $\theta = p/N$, the gauge coupling of the non-commutative
theory $g^2$ is equal to the 't~Hooft coupling of the commutative
theory $g_{com}^2N$. This relation holds when the couplings
are evaluated at the same scale. The 't~Hooft magnetic flux $m$
is a solution of $pm = 1 \mod N$, and is only defined modulo
$N$.  The radius of the non-commutative torus $R_{nc}$ is related to
that of the commutative torus $R_{com}$ by $R_{nc}= NR_{com}$.  The
momentum on the non-commutative torus is $\vec n/R_{nc}$,  which is
related to the 't~Hooft electric flux $\vec e$ by \elflux\ .

This last relation is crucial to us and can be understood in the
following way: the 't~Hooft electric fluxes of the $SU(N)/\MZ_N$ theory
are measured by operators associated with large gauge transformations on the torus.
Upon quotienting by the small gauge transformations, the large gauge
transformations generate a $\MZ_N \times \MZ_N$
global symmetry.  The small gauge transformations satisfy the same boundary
conditions as the fields,  which are given in \boundaryone, \boundarytwo, whereas the
large gauge transformations satisfy these boundary conditions
up to phase factors in the center of $SU(N)$.
In the presence of magnetic flux $m$ such that
$gcd(N,m)=1$, the
action of the large gauge transformations is equivalent, modulo small
gauge transformations, to translating the fields all the way around the
commutative torus.
Some details concerning large gauge transformations and
translations in the torus with twisted boundary conditions are discussed
in appendix~C.
The fact that the translation all the way around the torus is non-trivial
is due to the twisted boundary
conditions \footnote{${}^4$}{Note that the absence of zero modes in the presence of
twisted boundary conditions does not by itself break translation invariance,
since quantum mechanically,  there may still be a translationally invariant 
vacuum.}
which give fractional modes in the Fourier expansion of $A_{\mu}$.
While one can regard these modes as fractional momenta, a better
interpretation is as electric flux.  One can then define an integer
quantized momentum as
$$
{\cal P}_i =  P_i -   {\epsilon_{ij} e_j}\frac{m}{N R_{com}}.
$$
where $\vec P$ is the usual generator of translations,
$$
P_i = \int_{T^2 \times  \MR} d^3x {\rm tr} F_{i\mu}F_{0\mu}
$$
The electric fluxes (or fractional momenta) may
be reinterpreted as integer momenta upon rewriting the $SU(N)/\MZ_N$ theory as 
a non-commutative $U(1)$ gauge 
theory, due to the relation $R_{nc}=NR_{com}$.
Translations all the way around the commutative torus are equivalent
to translations of the
non-commutative fields by distances which are an integer multiples of
$2\pi R_{nc}/N$. For this reason condensation of electric flux in the
commutative theory via $\MZ_N \times \MZ_N$ symmetry breaking is
equivalent to spontaneous breaking of translation invariance in the
non-commutative description.  Of course it may also be viewed as spontaneous
breaking of translation invariance in the commutative theory if one takes
$\vec P$ for the generator of translations rather than $\vec{\cal P}$.

Before discussing symmetry breaking, there is a subtlety which we shall
address. Since we are considering the case in which two dimensions are
compactified on
a torus,  it might naively appear that the theory is effectively $D-2$
dimensional and that for $D \le 4$ quantum fluctuations should wash out any
symmetry breaking. As is well known,
there is
no spontaneous breaking of either continuous or discrete global symmetries
in theories with dimension less than $2$,  and no spontaneous breaking of
continuous symmetries \cite{Co} in theories with
dimension $2$ \footnote{${}^5$}{A possible exception arises in an
$N\rightarrow\infty$ limit.}.  Note that the $\MZ_N \times \MZ_N$ symmetry
is embedded in a continuous global symmetry generated by $\vec P$.
However, since the symmetry which is potentially broken
is a translation invariance in the compact directions,  one must proceed
with some care. In the
absence of twisted boundary conditions, translation invariance on the torus 
is not visible after
integrating out modes above the scale $1/R_{com}$.  However with
twisted boundary conditions,  the fractional momenta survive
below the scale $1/R_{com}$ and there is a global $\MZ_N \times \MZ_N$
symmetry in the dimensionally reduced theory. Furthermore,
the continuous symmetry in which this is embedded is no longer visible.
To see this explicitly in the $SU(N)/\MZ_N$ description,  it is useful to
write the degrees of freedom of the reduced theory in a gauge invariant
way.  All local gauge invariant operators, such as ${\rm tr F^2}$ are
periodic on the commutative torus and have integer momenta.
The fractional modes $A^{\mu}(\kappa)$ are associated to
the Wilson loops wrapping cycles of the torus (see \cite{vB1},\cite{vB2} and
appendix~C), which
are the creation operators for electric flux.
The leading term in the expansion of the path ordered exponential in
a Wilson loop is
proportional to a fractional momentum mode $A^{\mu}(\kappa)$.
Thus below the scale
$1/R_{com}$ the fundamental degrees of freedom are,
up to gauge transformations,
$SU(N)$ valued matrices $g_i$, from which the wrapped Wilson loops are 
built (see appendix C).  The index $i$ indicates a cycle of the torus.
These are the natural gauge invariant completions of the fractional modes $A_\mu(\kappa)$. 
Note that in the non-compact non-commutative case the one-loop effective action can be written naturally in terms of
open Wilson loop operators \cite{KRSY1},  \cite{KRSY2}, \cite{vR}, \cite{AL},
\cite{KKRS}, which are analogous to the wrapped 
Wilson loops introduced above.

A general mode with arbitrary fractional momenta and polarization is built
from traces of words written with $g_1$ and $g_2$. 
The $\MZ_N \times \MZ_N$ symmetry acts by multiplying the $g_i$ with phases
in the center of $SU(N)$.  Of course since $g_i$ are $SU(N)$ valued, an
arbitrary phase is not possible. Moreover there
is a cutoff on the fractional momentum arising from the fact that Wilson
loops annihilate modulo $N$; traces of sufficiently high
powers of $g_i$ (i.e. greater than $N$) can be written in terms of operators
built from traces with smaller powers of $g_i$. In this sense,
the dimensionally
reduced theory below the scale $1/R_{com}$
resembles a lattice discretization of
the non-commutative torus! Quantum fluctuations in two dimensions
present no obstruction
to spontaneously breaking the discrete $\MZ_N \times \MZ_N$ symmetry.

Let us now examine the phase structure of the four dimensional theory,
recalling that
the one loop dispersion relation for a photon propagating
along a cycle of the two-torus has the general form
$$
E^2 = \frac{1}{R_{nc}^2}\left(\vec \kappa^2 - g^2f\left(\frac{\vec e}{N}\right)\right) \gln\dispn. 
$$
The gauge coupling in \dispn\ should be evaluated at the scale $\kappa_i/R$.
The function $f(\frac{\vec e}{N})$ (see \light)
depends upon the polarization.
As discussed in the previous section, for transverse
polarization along the torus this function can lead to instabilities.
In these cases, $f$ is positive and
takes its largest
value when $e_i$ is smallest\footnote{${}^6$}{Recall that $f$ vanishes
when $e_i$ is zero,  since such modes correspond to
a decoupled sector.}. For small $\vec e$, $f\sim \frac{N^2}{\vec e ^2}$.

First let us consider the phase structure as a function of $R$ for
fixed $\Lambda_{QCD}$ and fixed $N$.  For $R_{com} \gg \frac{1}{\Lambda_{QCD}}$ one
expects the energies of electric fluxes to be controlled by
a confining linear
potential and there should then be no $\MZ_N \times \MZ_N$ symmetry
breaking via electric flux condensation.
At smaller radii,  such that the 't~Hooft coupling is small,  the dispersion
relation \dispn\ can become tachyonic for sufficiently large $N$,
leading to electric flux condensation.  However due to asymptotic
freedom,  the one loop effects of \dispn\ become small compared to the
kinetic term below a critical radius,  leading to
symmetry restoration.
Thus one has a rather unusual phase diagram,  in which symmetry breaking
occurs only at intermediate radii.  Only the critical point at
small radius can be studied perturbatively.  The phase transition at this 
point is apparently second order.

We emphasize that the kinetic term in \dispn\ is
a feature which arises only in the case of twisted boundary conditions,
since only in this case do wrapped Wilson loops carry momentum
(see appendix~C).  Thus at tree level,  the effect of the twisted
boundary conditions is
to {\it prevent} electric flux condensation.
This is very closely related
to phenomena arising in twisted Eguchi-Kawai models.  Such models
have been used to circumvent the $\MZ_N$ symmetry breaking
characteristic of the untwisted Eguchi-Kawai model. However,
we find that the one-loop correction can in fact lead to electric flux condensation.

For completeness, let us compare the phase structure of four dimensional 
$SU(N)/\MZ_N$
with twisted boundary conditions (magnetic flux) on a $T^2$  to that of the 
same
model without magnetic flux.  In the latter case,   one expects $\MZ_N 
\times \MZ_N$
symmetry breaking at small radius via arguments analogous to those of 
\cite{SY1}, \cite{SY2} in
the context of finite temperature deconfinement.  The difference in this 
case is that
there is no restoration of the symmetry as the radius goes to zero.

It is interesting to consider the phase structure in various large $N$
limits\footnote{$^7$}{Such a study has been performed for a scalar field model in \cite{MRW}.}.  
For simplicity we will consider the case in which
$\theta = \frac{1}{N}$,  or $m=N-1$. The physics of the large $N$
limit depends crucially on how one scales $R_{com}\Lambda_{QCD}$.
For instance,
keeping $R_{com}\Lambda_{QCD} = R_{nc}\Lambda_{QCD}/N$ and $m/N$ fixed 
corresponds to a conventional
planar 't~Hooft limit, whereas scaling $R_{com}\Lambda_{QCD}$
like $1/\sqrt{N}$ describes  a decompactification limit of the
non-commutative
theory \cite{G}\cite{AB}.  In the latter case the dimensionful non-commutative
parameter $\vartheta = 2 \pi \theta R_{nc}^2$ is held fixed while 
$R_{nc}\rightarrow\infty$
like $\sqrt{N}$. However, for a precise discussion the coupling in \dispn\ 
should be evaluated at the proper scale. As the running in the non-commutative
theory is non-standard in the infrared \cite{KT} and one must take into account possible
finite size complications we will not discuss this issue further here.  

When $N$ is large but finite, we expect 
a non-symmetric but stable vacuum, since it would be very surprising if 
$SU(N)/\MZ_N$ Yang-Mills theory
on a twisted torus had no stable vacuum.
Although it would be interesting to do so,
we will not attempt to find the stable vacuum here.
Note that the $N\rightarrow\infty$ limit with
$R_{com}\Lambda_{QCD} \sim 1/{\sqrt N}$ may be very singular,  and it is not 
clear
that the vacuum behaves smoothly in this limit.  String theory arguments 
suggest that the
theory on the non-commutative plane has no vacuum \cite{vR}, which
leads us to presume that there is no such smooth limit.

\chapter{Conclusions}
We have studied perturbative corrections to correlators for field
theories on the non-commutative torus. In the  case of the
non-commutative plane, these theories are known to show
UV/IR-mixing. Loop-corrections that are UV divergent in the
commutative limit are regulated by the non-commutativity. On the other hand
this regulator is removed for small external momenta and the divergencies
reappear.

In the compact case an important ingredient is Morita equivalence. It is
possible to parameterize the non-commutativity by a dimensionless parameter.
This parameterization is not unique but a $SL(2,\MZ)$ symmetry relates
different values. In the case of rational $\theta$ it is possible to
reformulate the theory on an ordinary, commutative space. Here, one would not
expect an exotic behavior such as UV/IR-mixing.

We have demonstrated that this puzzle can be explained by two observations:
first, as the commutative theory is formulated in terms of matrices, what has
been momentum in the non-commutative case now has to be split into a label for
matrix entries and commutative momentum. Second, as loop integrals are
replaced by sums these are subject to Poisson resummations that transform
oscillating phase factors into shifts in the sums over momenta.
In the rational case, these shifts affect only the matrix entry part and not
the commutative momentum part after the Morita map. Thus the commutative
theory does not show an unexpected dependence on external momenta.

But even if there is no new dependence on the commutative momenta, the
dependence on the matrix entry labels can be quite strong. In fact, for the
Moyal bracket interaction and also in the case of the gauge theory, the one
loop correction can render some of the modes perturbatively unstable just
as it has been known for the theory on the non-commutative plane. We have
discussed the nature of these instabilities and their relation to perturbative
$\MZ_N \times \MZ_N$ symmetry breaking in the gauge theory case.

As the twisted boundary conditions tie large gauge transformations to
translations along the cycles of the torus, a breakdown of the symmetry might
also be interpreted in terms of spontaneous breaking of translation
invariance due to electric flux condensation. It would be very interesting to
better understand the nature of this phenomenon and whether it persists in a
full non-perturbative quantum formulation of the theory.
\bigskip
\noindent{\bf Acknowledgments}
\nobreak
\noindent 
We would like to thank L. Alvarez-Gaum\'e, J. Barb\'on, 
W. Bietenholz, M. Garc{\'\i}a-P\'erez, C. G\'omez, A.~Gonz\'alez-Arroyo, C. P. 
Korthals Altes and D.~L\"ust for discussions. ZG thanks CERN for hospitality. 
We also would like to express our gratitude to EG for spaghetti with meatballs.
ZG is supported by the DFG under the Emmy Noether programme, grant ER301/1--2,
KL is supported by the DFG Schwerpunktprogramm SPP 1096.
RCH is supported by the DFG.

\vfill
\eject
\chapter{Appendix A: Integral representation of the Epstein Zeta Function}
In this appendix we review the analytic continuation of the Epstein zeta 
function over the complex $s$-plane using the integral representation in 
terms of Jacobi-theta functions \cite{M}.
For $Re(s)>1$ the Epstein zeta function is defined through the sum
$$
\zeta(s,\vec{b}) := \sum_{l\in \MZ^2}^{} {}^{'} [ ( \vec{l}+\vec{b} )^2 ]^{-s}\,.
$$
We assume that the components of $\vec{b}$ lie both in $(-{1\over 2}, {1\over 2} ]$. The prime
in the sum means that the origin is omitted in the case $\vec{b}=(0,0)$. 
We can represent this in integral form as
$$
\zeta(s,\vec{b}) = {\pi^s\over \Gamma(s)} \int_0^\infty \sum_{l\in\MZ_2} dt\,t^{s-1}\left[
e^{-\pi t (\vec{l}+\vec{b})^2} - \delta_{0,\vec{b}}\right]\,.
$$
In the next step we split the integral from $0$ to $1$ and from $1$ to $\infty$.
In the first integral we perform then a Poisson resummation 
$$
\sum_{n\in\MZ} e^{-a \pi n^2+ i 2 \pi n b } = {1\over \sqrt{a}} \sum_{l\in \MZ} e^{-{\pi\over a} (l+b)^2 }\,,
$$
and also change the integration variable $t\rightarrow 1/t$. We integrate the
zero-mode in the Poisson resummed expression explicitely and arrive at
$$\eqalignno{
\zeta(s,\vec{b}) = & {\pi^s\over \Gamma(s)} \left( {1\over s-1} - {\delta_{0,\vec{b}}\over s}
+\right.\cr
& \left.\int_1^{\infty}  dt \{ t^{s-1} [ \sum_{l\in\MZ_2}   e^{-\pi t (\vec{l}+\vec{b})^2}-
\delta_{0,\vec{b}} ] + t^{-s} [ \sum_{n\in\MZ_2} e^{-\pi t \vec{n}^2 + i 2 \pi \vec{b}\cdot\vec{n} }
-1]\}\right)\,.}
$$
Finally the sums can be written as Products of the Jacobi-Theta function
$$
\vartheta(z,\tau) := \sum_{n\in\MZ} e^{i\pi \tau n^2 + i 2 \pi n z}\,,
$$
which gives the form used in equ. \defzeta.

The Epstein zeta function has the following properties.
It is meromorphic on the complex $s$-plane with a simple pole at $s=1$ and
residue $\pi$.
The behavior at the origin is given by $\zeta(0,\vec{b}) = - \delta_{0,\vec{b}}$ and
more generally $\zeta(-n,\vec{b})=0$ for $n\in \MN$.

\chapter{Appendix B: Polarization tensor in the NC Torus}
In this Appendix we will derive expression \pol-\seriesb\
for the non-planar part of the polarization tensor.
Using the non-commutative Feynman rules, the non-planar 
contribution to the polarization tensor is given by
$$
\Pi^{NP}_{\mu \nu} = {g^2 \over (2 \pi R)^2} 
\sum_{\vec{l} \in\MZ} \int
{d^d k \over (2 \pi)^d} {{\rm cos} (2 \pi \theta n_i l_i \epsilon_{ij})
\over \big(k^2 + {\vec{l}^2 \over R^2}\big)\big((k-p)^2+
{(\vec{n}-\vec{l})^2 \over R^2}\big)}
f_{\mu \nu} (K_\rho,P_\sigma) \, ,
$$
where $P_\mu=(p_\alpha,n_i/R)$ is the inflowing momentum, 
$K_\mu=(k_\alpha, l_i/R)$ denotes the loop momentum
and $f_{\mu \nu}(L_\rho,P_\sigma)$ is
$$
4d \, K_\mu K_\nu - 2d \, (P_\mu K_\nu + P_\nu K_\mu) 
+(d-4)\, P_\mu P_\nu 
- g_{\mu \nu}\, \big( 2d \,K^2 -(4d +2)\, K\!\cdot\!P+(2d-3)\,P^2 \big) \, .
$$

In order to evaluate the previous expression we introduce Feynman 
and Schwinger parameters, which allow us to easily integrate the internal 
momentum along the non-compact directions, $k_\alpha$. We expand then the
cosine in terms of exponentials and Poisson resum, with the result
$$
\Pi^{NP}_{\mu \nu} =  
{g^2 \over 2 \, (4 \pi)^{{d \over 2}+1}} \sum_{\epsilon=
\pm 1} \sum_{\vec{l} \in\MZ^2} \,
\int_0^{\infty}\, {d \alpha \over \alpha^{d \over 2}}\, \int_0^1\, dx 
\, e^{-2 \pi i\epsilon x P\!\cdot\!L} \,
e^{- \alpha P^2 x(1-x) -{\pi^2 \over \alpha} L^2} 
\, f_{\mu \nu}(P_\rho,L_\sigma) \, , \gln\pol
$$
where we have set for simplicity $R=1$. We recall that 
$L_\mu=\big( 0,l_i - \theta \epsilon_{ij}n_j \big)$. The polynomial 
$f_{\mu \nu}$ after these manipulations reads
$$\eqalignno{
{f'}_{\mu \nu}(P_\rho,& \, L_\sigma)  = \;- {2d \pi^2 \over \alpha^2}
\left[2  L_\mu L_\nu -  \, g_{\mu \nu}
\, \left( L^2- {d \over  2 \pi^2}\alpha  \right) \right] \, - \cr
&-{i \pi \epsilon \over \alpha} \left[ 2d \, ( 2x-1) \, 
\big( P_\mu L_\nu + P_\nu L_\mu) +
g_{\mu \nu} \, (4d (1-x)+2) \, P\!\cdot\! L \right]\, +\cr
&+ \big( 4d \, x(1-x)-d +4 \big)P_\mu P_\nu
-g_{\mu \nu} (2d \, (x-1)^2 -2 \, x -3) P^2 \, . \egln\expra}
$$  

This complicated expression can be simplified by making use of the
properties of Bessel functions. Let us start by concentrating our 
attention on the first line in the rhs of \expra. Since it carries
negative powers of the Schwinger parameter $\alpha$, it can provide 
very large contributions from the UV region 
$\alpha \rightarrow 0$. Notice that the second term inside the square
bracket is proportional to $g_{\mu \nu}$ and thus could threaten
the Ward identities. However a cancellation takes place, as 
can be seen from the following relations
$$\eqalignno{
\int_0^{\infty} {d \alpha \over \alpha^{{d \over 2}+2}} \, 
e^{- z^2 \alpha -{1 \over \alpha}} 
\Big( 1  - {d \over 2 } \alpha \Big) & = 
2 z^{{d \over 2}+1}
\Big( K_{{d \over 2}+1} (2 z)-
{d \over 2z}  K_{d \over 2} (2z) \Big)= \cr
&=2 z^{{d \over 2}+1}  K_{{d \over 2}-1}(2z) =
z^2 \int_0^{\infty} {d \alpha \over \alpha^{d \over 2}}  \, 
e^{- z^2 \alpha-{1 \over \alpha}} \; .}
$$
This relation applies to \pol-\expra\ by setting $z^2=\pi^2 L^2 P^2 x(1-x)$.
It implies that we can substitute
$$
- {2d  \over \alpha^2} g_{\mu \nu}
\, \left(\pi^2 L^2- {d \over  2}\alpha  \right)
\rightarrow -2d P^2 x(1-x) \, 
$$
in ${f'}_{\mu \nu}$, allowing us to rewrite it in the form
$$\eqalignno{
{f''}_{\mu \nu}(P_\rho,L_\sigma) =&  \; -{4d \pi^2 \over \alpha^2} \, 
L_\mu L_\nu 
\, - {2 \pi i d \epsilon  \over \alpha} (2x -1) \left[
P_\mu L_\nu + P_\nu L_\mu -g_{\mu \nu}P\!\cdot\!L \right] \cr
& -
(P_\mu P_\nu-g_{\mu \nu}P^2) \big( \, 4d \, x\,(1-x)-d+4\, \big)\, + \cr
& + g_{\mu \nu} (d+1) \Big[(2x-1)P^2 -{2 i \epsilon \pi \over \alpha} 
P\!\cdot\!L \Big]\, . \egln\exptwo}
$$
In order to proceed we will show that the following equation holds
for $r \in \MZ$ 
$$\eqalignno{
\int_0^1 dx \, e^{2 \pi i r x}& (2x-1) \int_0^\infty
{d \alpha \over \alpha^{n}} e^{- \alpha z^2 x(1-x)-{1 \over \alpha}}= \cr
& =-{2 \pi i r \over z^2} \int_0^1 dx \, e^{2 \pi i r x}
\int_0^\infty {d \alpha \over \alpha^{n+1}} 
e^{- \alpha z^2 x(1-x)-{1 \over \alpha}} \, . \egln\rel}
$$
The lhs side of this equation is
$$\eqalignno{
\int_0^1 dx \,& e^{2 \pi i r x} (2x-1) \int_0^\infty
{d \alpha \over \alpha^{n}} e^{- \alpha z^2 x(1-x)-{1 \over \alpha}}= \cr
& = 2 \int_0^1 dx \, e^{2 \pi i r x} (2x-1) \big( z \sqrt{x(1-x)} \big)^{n-1}
K_{n-1}(2 z \sqrt{x(1-x)}) = \cr
& = {2 \over z^2} \int_0^1 dx \, e^{2 \pi i r x} \partial_x \Big[
\big( z \sqrt{x(1-x)} \big)^{n} K_{n}(2 z \sqrt{x(1-x)}) \Big]= \, , }
$$
where we have used that $\partial_y (y^{n} K_{n}(y))=-y^{n} K_{n-1}(y)$. 
Provided $r$ is an integer, integrating by parts we obtain the desired
relation
$$\eqalignno{
&= -{4 \pi i r \over z^2}  \int_0^1 dx \, e^{2 \pi i r x}
\big( z \sqrt{x(1-x)} \big)^{n} K_{n}(2 z \sqrt{x(1-x)}) = \cr
&=-{2 \pi i r \over z^2} \int_0^1 dx \, e^{2 \pi i r x}
\int_0^\infty {d \alpha \over \alpha^{n+1}} 
e^{- \alpha z^2 x(1-x)-{1 \over \alpha}} \, .}
$$
Setting $r=\epsilon P\!\cdot\!L$ and $z^2 = \pi^2 L^2 P^2$, the equality \rel\
implies that we can replace 
$$
(2x-1) \rightarrow {2 i \epsilon \pi \over \alpha} 
{ P \!\cdot\!L \over P^2} \, 
$$
in the involved expression of ${f''}_{\mu \nu}$.
It is clear then that the last line in \exptwo\ cancels, and
${f''}_{\mu \nu}$ reduces to
$$\eqalignno{
h_{\mu \nu}(P_\rho,L_\sigma) =&  \;- {4d \,\pi^2 \over \alpha^2} 
\left[  L_\mu L_\nu 
\, - \, \big( P_\mu L_\nu + P_\nu L_\mu - g_{\mu \nu} P\!\cdot\!L \big)
\, {P\!\cdot\!L \over P^2} \, \right] \, 
-  \cr
& - \big( \, P_\mu P_\nu - g_{\mu \nu} P^2 \big) \, 
\big(\, 4d \,x\, (1-x)-d+4 \, \big)  \, .}
$$

\chapter{Appendix C: Translations and large gauge transformations on a 
twisted torus}
In this appendix we review some general properties of translations
and large gauge transformations on a twisted two-torus. We
consider an $SU(N)/\MZ_N$ gauge theory on a torus with twisted
boundary conditions corresponding to magnetic flux $m$.  The
$SU(N)$ valued twist matrices associated to the cycles of the
torus are $U_1$ and $U_2$ which satisfy
$$
U_1 U_2=U_2 U_1 \exp(-2\pi i \frac{m}{N}),
$$
and may be chosen to be constant.
Translating the gauge fields all the way around the cycles of the torus is 
equivalent to acting
with the gauge transformations $U_1$ and $U_2$.

Physical states are invariant under Small gauge transformations. These
are continously connected to the identity and are described by
$U_s(x)$ satisfying 
$$
U_s(x^i +2\pi R^i) = U_iU_s(x)U_i^{\dagger}.
\gln\small
$$
However, there are more transformations that preserve the boundary
conditions for the connection. In addition to the small
transformations there are transformations for which $U_s(x)$ obeys
\small\ only up to a phase. This phase has to be an $N$th root of
unity in order for $U_s$ to be in $SU(N)$:
$$
U_l(x^i +2\pi R^i) = U_iU_l(x)U_i^{\dagger}\exp(2\pi i 
\frac{n_i}{N})
\gln\large
$$
Note that the twists $U_1$ and $U_2$ are of this type.
For instance
$$
U_2 = U_1U_2U_1^{\dagger}\exp(2\pi i m/N)
$$
There is a $\MZ_N \times \MZ_N$ global symmetry associated to the quotient 
of large
gauge transformations by small ones.  For the case $gcd(N,m) =1$,  which is 
dual
a non-commutative $U(1)$ theory,  this symmetry is completely generated by 
the twist
matrices,  whereas in other cases the twist matrices only generate a 
subgroup.

States that transform non-trivially under large gauge transformations are the 't~Hooft electric fluxes.
The associated creation operators are wrapped Wilson loops.  Because of the 
relation
between the large gauge transformations and the twists, or translations all 
the way around
the torus,  one expects such operators to have momentum which is fractional 
in units of 1/N.
For instance the Wilson loops which wrap the $i$'th cycle of the torus $e$ times are 
given by
$$
W_{i,e} = {\rm tr} P\left(e^{i\oint_{e S_i} A_i}\right){U_i^{\dagger}}^e
$$
where the twist matrix in this expression is necessary for the operator to 
be invariant under
small gauge transformations \cite{vB1}, \cite{vB2}.
Note that this can also be written as
$$
W_{i,e} = {\rm tr} g_i^e
$$
where
$$
g_i =  P\left(e^{i\oint_{S_i} A_i}\right)U_i^{\dagger} .
$$
Upon translating all the way around the torus in the transverse
direction,  this operator picks up the phase $\exp(2\pi i m e/N)$.
In an effective action for the fractional momentum modes, these Wilson loop 
operators are
the natural gauge invariant completions of gauge field modes with fractional 
momenta.
For example, expanding the exponential gives
$$
{\rm tr} g_1^{e_1} =  2\pi \kappa_2 R_{com} A_1(\kappa_2) + \dots
$$

In a non-abelian gauge theory,  the translation generator
$p_j=\int {tr}F_{0i}\partial_jA_i$ is not gauge invariant. However
the gauge invariant version $P_j = \int {tr}F_{0i}F_{ji}$
generates exactly the same action on physical states, since the
action of the two generators differs by a small gauge
transformation.  For instance,  the infinitesimal translation
$\exp(i\epsilon p_j)$ differs from $\exp(i\epsilon P_j)$ by a
gauge transformation with the gauge parameter $\alpha = \epsilon
A_j$. A translation all the way around the torus is generated by
$\exp( 2\pi i R P_j)$,  which in light of our previous discussion is
a large gauge transformation. Thus when there is a non-zero
electric flux,  the momentum $P_j$ may be fractional in units of
$1/N$.  The fractional momenta are better interpreted as electric
fluxes. One can define a momentum with integer eigenvalues by
$$
{ {\cal P}_j} \equiv { P}_j -
m\epsilon_{ji}{ e}_i/N.
$$

\closebib
\chapter{References}
\bigskip
\parindent =2cm 
\parskip=10 pt plus 5pt
\paper{AB}{Alvarez-Gaume, L. and Barbon, J. L. F.}{Morita duality and large-N limits}{Nucl. Phys.}{B623}{2002}{165--200}{{\tt hep-th/0109176} }{Cited: 3 21 }
\eprint{AL}{Armoni, Adi and Lopez, Esperanza}{UV/IR mixing via closed strings and tachyonic                  instabilities}{hep-th/0110113}{}{Cited: 2 20 }
\paper{BGNV}{Bassetto, A. and Griguolo, L. and Nardelli, G. and Vian, F.                  }{On the unitarity of quantum gauge theories on                  noncommutative spaces}{JHEP}{07}{2001}{008}{{\tt hep-th/0105257} }{Cited: 2 11 }
\buch{C}{Connes, Alain}{Noncommutative Geometry}{1994}{Academic Press}{}{}{Cited: 2 }
\paper{CMS}{Chu, Chong-Sun and Madore, John and Steinacker, Harold}{Scaling limits of the fuzzy sphere at one loop}{JHEP}{08}{2001}{038}{{\tt hep-th/0106205} }{Cited: 2 }
\paper{Co}{Coleman, Sidney R.}{There are no Goldstone bosons in two-dimensions}{Commun. Math. Phys.}{31}{1973}{259--264}{{\tt } }{Cited: 19 }
\paper{DFR}{Doplicher, S. and Fredenhagen, K. and Roberts, J. E.}{The Quantum structure of space-time at the Planck scale and                  quantum fields}{Commun. Math. Phys.}{172}{1995}{187--220}{{\tt } }{Cited: 2 }
\paper{DN}{Douglas, Michael R. and Nekrasov, Nikita A.}{Noncommutative field theory}{Rev. Mod. Phys.}{73}{2002}{977--1029}{{\tt hep-th/0106048} }{Cited: 2 }
\paper{F}{Filk, T.}{Divergencies in a field theory on quantum space}{Phys. Lett.}{B376}{1996}{53--58}{{\tt } }{Cited: 6 }
\eprint{G}{Guralnik, Zachary}{Strong coupling phenomena on the noncommutative plane}{hep-th/0109079}{}{Cited: 21 }
\paper{GK}{Gonzalez-Arroyo, A. and Korthals Altes, C. P.}{Reduced Model For Large N Continuum Field Theories}{Phys. Lett.}{B131}{1983}{396}{{\tt } }{Cited: 2 6 }
\paper{GMW}{Gomis, Jaume and Mehen, Thomas and Wise, Mark B.}{Quantum field theories with compact noncommutative extra                  dimensions}{JHEP}{08}{2000}{029}{{\tt hep-th/0006160} }{Cited: 3 }
\paper{GP}{Griguolo, Luca and Pietroni, Massimo}{Wilsonian renormalization group and the non-commutative                  IR/UV  connection}{JHEP}{05}{2001}{032}{{\tt hep-th/0104217} }{Cited: 2 }
\eprint{GT}{Zachary Guralnik and Jan Troost}{Aspects of gauge theory on commutative and noncommutative                  tori}{hep-th/0103168}{}{Cited: 3 }
\eprint{H}{Helling, Robert C.}{A remark on field theories on the non-commutative torus}{hep-th/0111077}{}{Cited: 3 }
\eprint{Ha}{Hayakawa, M.}{Perturbative ultraviolet and infrared dynamics of                  noncommutative  quantum field theory}{hep-th/0009098}{}{Cited: 2 14 }
\eprint{KKRS}{Kiem, Young-jai and Kim, Yeon-jung and Ryou, Cheol and                  Sato, Haru-Tada}{One-loop noncommutative U(1) gauge theory from bosonic                  worldline  approach}{hep-th/0112176}{}{Cited: 2 20 }
\paper{KRSY1}{Kiem, Youngjai and Rey, Soo-Jong and Sato, Haru-Tada and                  Yee, Jung-Tay}{Open Wilson lines and generalized star product in                  nocommutative scalar  field theories}{Phys. Rev.}{D65}{2002}{026002}{{\tt hep-th/0106121} }{Cited: 2 20 }
\paper{KRSY2}{Kiem, Youngjai and Rey, Soo-Jong and Sato, Haru-Tada and                  Yee, Jung-Tay}{Anatomy of one-loop effective action in noncommutative                  scalar field  theories}{Eur. Phys. J.}{C22}{2002}{757--770}{{\tt hep-th/0107106} }{Cited: 2 20 }
\paper{KT}{Khoze, Valentin V. and Travaglini, Gabriele}{Wilsonian effective actions and the IR/UV mixing in                  noncommutative  gauge theories}{JHEP}{01}{2001}{026}{{\tt hep-th/0011218} }{Cited: 21 }
\paper{KW}{Krajewski, Thomas and Wulkenhaar, Raimar}{Perturbative quantum gauge fields on the noncommutative                  torus}{Int. J. Mod. Phys.}{A15}{2000}{1011--1030}{{\tt hep-th/9903187} }{Cited: 2 }
\paper{LLT1}{Landsteiner, Karl and Lopez, Esperanza and Tytgat, Michel                  H. G.}{Excitations in hot non-commutative theories}{JHEP}{09}{2000}{027}{{\tt hep-th/0006210} }{Cited: 2 }
\paper{LLT2}{Landsteiner, Karl and Lopez, Esperanza and Tytgat, Michel                  H. G.}{Instability of non-commutative SYM theories at finite                  temperature}{JHEP}{06}{2001}{055}{{\tt hep-th/0104133} }{Cited: 2 11 }
\buch{M}{David Mumford}{Tata Lectures on Theta I, (chap. I.16)}{1983}{Birkh\"auser Boston-Basel-Stuttgart}{}{}{Cited: 23 }
\eprint{MRW}{Mandal, Gautam and Rey, Soo-Jong and Wadia, Spenta R.}{Quantum aspects of GMS solutions of noncommutative field                  theory and  large N limit of matrix models}{hep-th/0111059}{}{Cited: 21 }
\paper{MST}{Matusis, Alec and Susskind, Leonard and Toumbas, Nicolaos}{The IR/UV connection in the non-commutative gauge                  theories}{JHEP}{12}{2000}{002}{{\tt hep-th/0002075} }{Cited: 2 14 }
\paper{MvRS}{Minwalla, Shiraz and Van Raamsdonk, Mark and Seiberg,                  Nathan}{Noncommutative perturbative dynamics}{JHEP}{02}{2000}{020}{{\tt hep-th/9912072} }{Cited: 2 2 7 7 13 16 }
\eprint{PS}{Peskin, Michael E. and Schroeder, D. V.}{An Introduction to quantum field theory}{Reading, USA: Addison-Wesley (1995) 842 p}{}{Cited: 13 }
\paper{RR}{Ruiz, F. Ruiz}{Gauge-fixing independence of IR divergences in non-                  commutative U(1),  perturbative tachyonic instabilities and                  supersymmetry}{Phys. Lett.}{B502}{2001}{274--278}{{\tt hep-th/0012171} }{Cited: 2 11 }
\paper{S}{Saraikin, Kirill}{Comments on the Morita equivalence}{J. Exp. Theor. Phys.}{91}{2000}{653--657}{{\tt hep-th/0005138} }{Cited: 3 5 }
\paper{Sn}{Snyder, Hartland S.}{Quantized space-time}{Phys. Rev.}{71}{1947}{38--41}{{\tt } }{Cited: 2 }
\paper{SY1}{Svetitsky, Benjamin and Yaffe, Laurence G.}{Critical behaviour at finite temperature confinement transitions}{Nucl. Phys.}{B210}{1982}{423}{{\tt } }{Cited: 18 21 }
\paper{SY2}{Yaffe, L. G. and Svetitsky, B.}{First order phase transitions in the $SU(3)$ gauge theory at finite temperature}{Phys. Rev.}{D26}{1982}{963}{{\tt } }{Cited: 18 21 }
\paper{SZ}{Schwarz, Albert}{Morita equivalence and duality}{Nucl. Phys.}{B534}{1998}{720--738}{{\tt hep-th/9805034} }{Cited: 2 }
\eprint{vB1}{van Baal, Pierre}{QCD in a finite volume}{hep-ph/0008206}{}{Cited: 5 19 27 }
\eprint{vB2}{van Baal, Petrus Jacobus}{Twisted boundary conditions: A nonperturbative probe for                  pure nonabelian gauge theories}{PhD thesis, INIS-mf-9631}{}{Cited: 19 27 }
\paper{vR}{Van Raamsdonk, Mark}{The meaning of infrared singularities in noncommutative                  gauge theories}{JHEP}{11}{2001}{006}{{\tt hep-th/0110093} }{Cited: 2 20 21 }

\bye